\documentclass[lettersize,journal]{IEEEtran}
\usepackage{amsmath,amsfonts}
\usepackage{algorithmic}
\usepackage{algorithm}
\usepackage{array}

\usepackage{textcomp}
\usepackage{stfloats}
\usepackage{url}
\usepackage{verbatim}
\usepackage{cite}

\usepackage{tabularx}
\usepackage{pifont}
\usepackage{booktabs}
\usepackage{multirow}
\usepackage{graphicx}
\usepackage{subcaption}
\usepackage{todonotes}

\newcolumntype{Y}{>{\centering\arraybackslash}X}
\newcommand{\cmark}{\textcolor{green!60!black}{\ding{51}}}
\newcommand{\xmark}{\textcolor{red!70!black}{\ding{55}}}

\usepackage[table,xcdraw]{xcolor}

\begin{document}

\title{EdgeCIM: A Hardware--Software Co-Design for CIM-Based Acceleration of Small Language Models}

\author{
    Jinane Bazzi\textsuperscript{*}, Mariam Rakka\textsuperscript{*}, Fadi Kurdahi, Mohammed E. Fouda and Ahmed Eltawil\\
    \textsuperscript{*}Equal contribution
    \thanks{J. Bazzi and A. Eltawil are with King Abdullah University of Science and Technology, Thuwal, Saudi Arabia.}%
    \thanks{M. Rakka and F. Kurdahi are with the University of California Irvine, Irvine, CA, USA.}
    \thanks{M. Fouda is with Rain Neuromorphics Inc., San Francisco, CA, USA. Email: foudam@uci.edu}%
    \thanks{This work was supported by the King Abdullah University of Science and Technology through the Competitive Research Grant program
under grant URF/1/4704-01-01.}
}




\maketitle

\begin{abstract}
The growing demand for deploying Small Language Models (SLMs) on edge devices, including laptops, smartphones, {automotive and robotic computing platforms, and high-performance embedded systems}, has exposed fundamental inefficiencies in existing accelerators. While GPUs handle prefill workloads efficiently, the autoregressive decoding phase is dominated by GEMV operations that are inherently memory-bound, resulting in poor utilization and prohibitive energy costs at the edge. In this work, we present EdgeCIM, a hardware--software
co-design framework for accelerating autoregressive
decoding of SLMs on edge devices in latency-critical
applications. At its core is a CIM macro, implemented in 65~nm, coupled with a tile-based mapping strategy that balances pipeline stages, maximizing parallelism while alleviating DRAM bandwidth bottlenecks. Our simulator enables design space exploration of SLMs
up to 4B parameters, identifying hardware configurations
that optimize latency and energy according to a
configurable objective function. {On LLaMA3.2-3B (INT4), EdgeCIM achieves 1.76$\times$ and 3.21$\times$ higher throughput than NVIDIA's Jetson AGX Orin and Orin Nano Super operating in MAXN mode, respectively, and 7.07$\times$ higher throughput than Qualcomm's Snapdragon X Elite.} Extensive benchmarks on TinyLLaMA-1.1B, LLaMA3.2 (1B, 3B), Phi-3.5-mini-3.8B, Qwen2.5 (0.5B, 1.5B, 3B), SmolLM2-1.7B, SmolLM3-3B, and Qwen3 (0.6B, 1.7B, 4B) reveal that our accelerator, at INT4 precision, achieves on average 336.4 tokens/s and 173 tokens/J. {These results establish EdgeCIM as a compelling solution towards real-time, energy-efficient, latency-critical SLM inference at the edge.}
\end{abstract}


\section{Introduction}
 Language Models (LMs) have transformed Natural Language Processing (NLP), establishing new benchmarks in text generation, translation, summarization, and conversational AI \cite{brown2020language,devlin2019bert,vaswani2017attention}. These models, built on the transformer architecture, achieve remarkable accuracy but demand enormous computational and memory resources. While the earliest deployment of LMs was restricted to datacenters running on clusters of Graphics Processing Units (GPUs) and Tensor Processing Units (TPUs) \cite{jouppi2017datacenter}, the rapid rise of agentic AI systems and the demand for interactive, privacy-preserving applications have shifted attention towards a new paradigm: running Small LMs (SLMs) closer to the user, on edge devices such as laptops, smartphones, automotive and robotic computing platforms, and high-performance embedded systems \cite{tambe2021edgebert, hu2023llamacpp}. 

Decoder-only architectures, typified by autoregressive models such as GPT \cite{radford2018improving} and LLaMA \cite{touvron2023llama}, are a popular choice for this new paradigm. Their token-by-token decoding aligns naturally with interactive use cases like translation, voice assistants, and contextual dialogue systems. Decoder-only inference supports real-time streaming of outputs, which is an essential requirement in low-latency and user-facing applications. However, supporting such interactive applications on resource-constrained edge devices requires efficient execution of the autoregressive inference process. Specifically, decoder-only SLM inference can be divided into a General Matrix-Matrix multiplication (GEMM)-heavy prefill phase and a General Matrix-Vector multiplication (GEMV)-dominated decoding phase. For interactive edge applications, where inference typically operates at a batch size of one, recent profiling of LLaMA inference on CPUs and NPUs confirms that decoding dominates runtime, often consuming more than 70\% of the total inference time \cite{hu2023llamacpp}, indicating a need to accelerate the decoding phase further.

Compute-in-Memory (CIM) architectures provide an efficient substrate for GEMV operations by reducing transfers between memory arrays and arithmetic units~\cite{verma2019imc}. For transformers, multiple recent CIM accelerator designs have specifically targeted the self-attention mechanism \cite{sridharan2023xformer,yang2020retransformer}. Designs such as X-Former \cite{sridharan2023xformer} and TranCIM \cite{tu2023trancim} achieve SOTA results on encoder-style models such as BERT, but their evaluation neglects the unique characteristics of decoder-only inference: none of these works a) address the GEMV-dominated nature of autoregressive decoding in edge scenarios nor b) optimize for the end-to-end decoding phase where projection and linear stages of inference play an equally important role as attention.

In fact, the use of CIM primitives alone does not resolve the system-level bottlenecks of edge-scale autoregressive decoding. Although these architectures can functionally execute Transformer kernels and, in some cases, masked attention, adapting their mappings to batch-size-one SLM decoding under constrained external-memory bandwidth requires more than changing the attention mask. Their dataflows would need to be reorganized around single-token GEMVs, continuous model-weight streaming, a growing KV-cache, limited inter-token reuse, and stage-dependent resource sharing across projection, attention, output projection, and Feed-Forward Network (FFN). While SRAM-CIM arrays, hierarchical tiles, reduction trees, buffers, and pipelined data movement are well-established components of prior DNN and Transformer accelerators~\cite{shafiee2016isaac,ankit2019puma,sridharan2023xformer,tu2023trancim}, their joint optimization under these decoding characteristics presents distinct architectural challenges. At each generation step, only one new token is available, the projection and feed-forward layers reduce to large GEMVs with negligible inter-token reuse, and attention repeatedly accesses a KV-cache that grows with the generated sequence. Therefore, increasing the number of simultaneously operating CIM tiles can reduce compute time while proportionally increasing the weight partition that must be delivered from off-chip memory before that computation can proceed. This creates a decode-specific coupling among instantaneous tile parallelism, LPDDR bandwidth, partition size, and pipeline utilization that is not addressed by merely mapping individual Transformer kernels to CIM arrays.

To address these challenges, EdgeCIM introduces a unified end-to-end mapping and a bandwidth-aware active-tile organization. The total number of tiles determines the available on-chip compute capacity, whereas an independently selected subset of active tiles determines the
instantaneous execution and transfer width. Inactive tiles preload
subsequent weight partitions while the active subset computes, allowing
the framework to control the demand placed on off-chip memory without
fixing the physical architecture to the same degree of parallelism.
The optimal active-tile count, cluster organization, processing element granularity,
and interconnect widths are selected jointly for each SLM and precision
through design space exploration across the {end-to-end decoding process}. The key contributions of this work are:
\begin{itemize}
    \item {We characterize the end-to-end, batch-size-one decoding
    pipeline of decoder-only SLMs and formulate its mapping onto a
    digital SRAM-CIM fabric, including projection, KV-cache attention,
    output projection, gated feed-forward layers, normalization,
    activation, quantization, and Softmax.}

    \item {We introduce a bandwidth-aware active-tile execution
    organization that decouples the total instantiated CIM capacity
    from the number of tiles operating concurrently. By tuning the
    active-tile subset, EdgeCIM jointly controls instantaneous compute
    parallelism, streamed weight-partition size, and the overlap of
    DRAM transfers with CIM execution.}

    \item {We develop stage-specific mappings that reconfigure the same
    hierarchical fabric across head-parallel projection, block-wise
    KV-cache attention, chip-wide output-projection reduction, and
    concurrent gate/up feed-forward computation, rather than assigning
    fixed hardware engines to isolated Transformer kernels.}

    \item {We jointly explore cluster count, total and active tile
    counts, processing element granularity, precision, and hierarchical bus widths over
    approximately 3.1$\times$10\textsuperscript{6} candidate configurations as part of the design space exploration. This
    exposes model-dependent compute-bandwidth trade-offs that are not
    captured by fixed CIM architectures.}

    \item {We evaluate complete autoregressive decoding for SLMs up to
    4B parameters and compare throughput, energy efficiency, and memory-bandwidth requirements against commercial edge
    GPUs and NPUs.}
\end{itemize}

The remainder of this paper is organized as follows: Section~\ref{sec:background} provides background on small language models, compute-in-memory technologies, and SLM quantization. Section~\ref{sec:methodology} presents the proposed EdgeCIM framework, including its hardware architecture and dataflow mapping. Section~\ref{sec:implementation} describes the hardware--software co-optimization process. Section~\ref{sec:results} presents the experimental results and comparisons with existing accelerators. Finally, Section~\ref{sec:conclusion} concludes the paper.

\begin{figure}[!b]
\centering
\includegraphics[width=1\columnwidth]{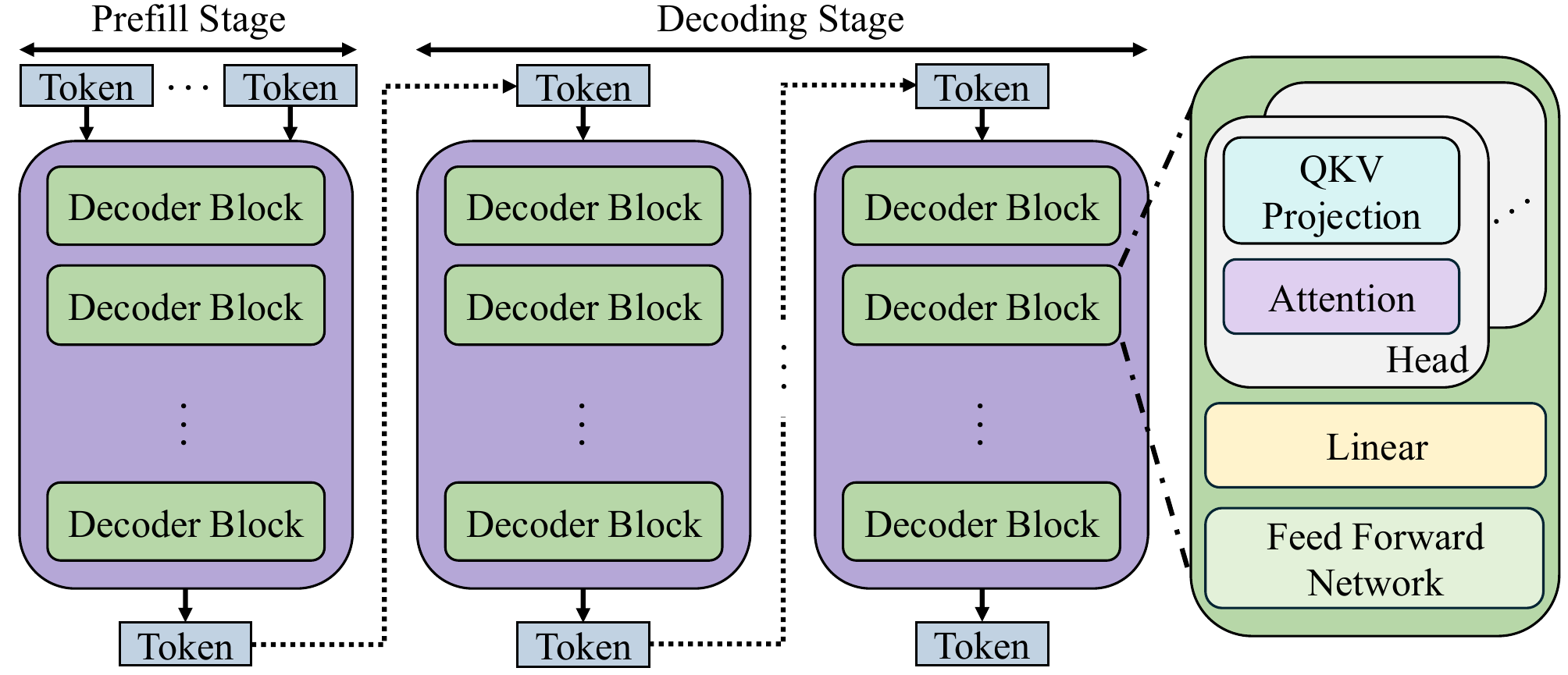}
\caption{Inference process in decoder-only SLMs.}
\label{decoder_archi}
\end{figure}

\section{Background}
\label{sec:background}
\subsection{Small Language Models}
SLMs have emerged as practical tools for text generation, summarization, and translation on resource-constrained platforms. They rely on attention mechanisms to capture long-range dependencies while operating within limited compute and memory budgets. Most SLMs adopt a decoder-only architecture, consisting of an embedding layer, stacked decoder blocks with Multi-Head Attention (MHA), Feed-Forward Networks (FFNs), normalization layers, and linear projections. MHA leverages $QKV$ projections, allowing the model to assign contextual importance to different tokens through the scaled dot-product attention mechanism. Recent variants such as Grouped-Query Attention (GQA) further improve memory efficiency by allowing multiple queries to share the same keys and values.

\begin{figure}[!t]
\centering
{\includegraphics[width=1\columnwidth]%\columnwidth]
{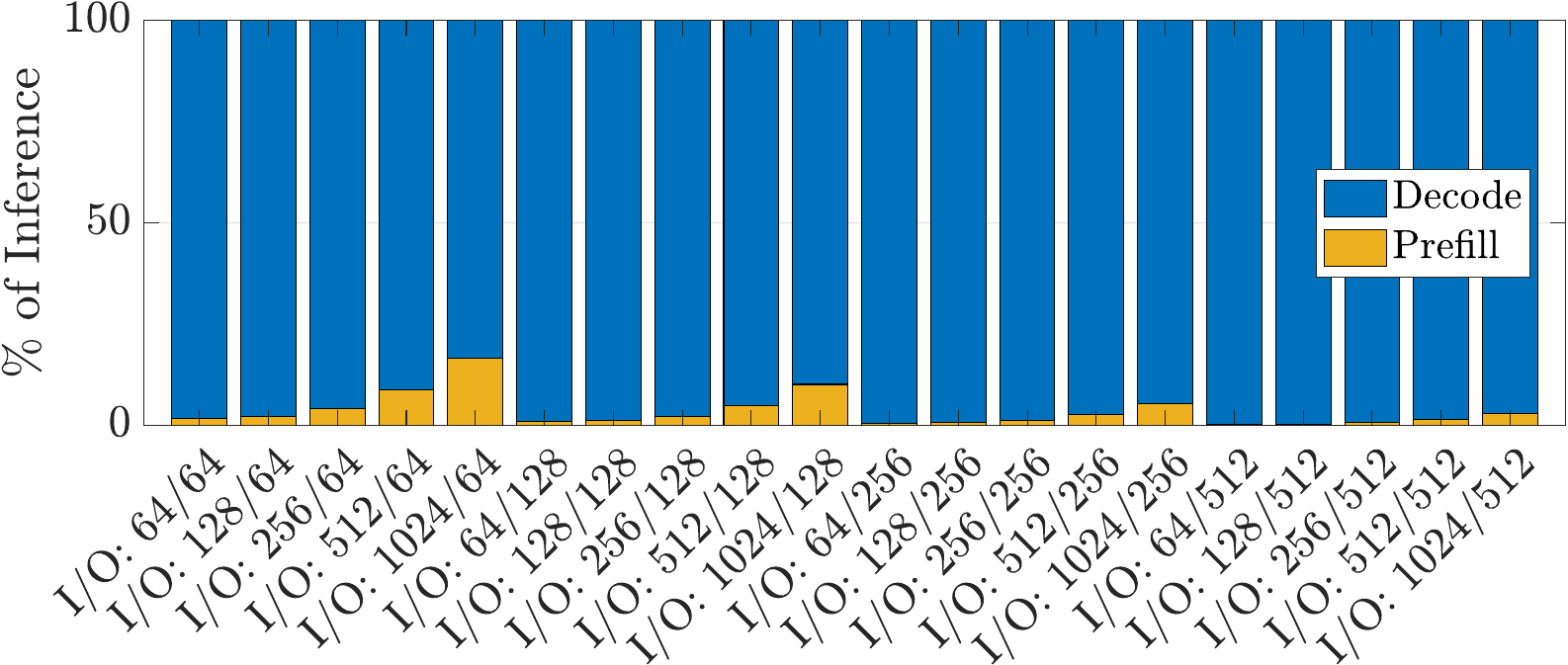}}
\caption{Profiling of LLaMA 3.2-1B on the NVIDIA Jetson AGX Orin GPU for different numbers of input ($I$) and output ($O$) tokens.}
\label{profiling}
\end{figure}

The inference process of autoregressive SLMs (Fig.~\ref{decoder_archi}) is divided into two stages: \textit{prefill}, which processes the input sequence and populates the KV-cache for all prompt tokens, and \textit{decoding}, which generates tokens one at a time by attending to the cached KV and the newly computed KV of the current token. Prefill exhibits high parallelism and is well handled by systolic-array-based accelerators, whereas decoding is dominated by GEMVs, which are memory-bound and underutilize conventional SIMD or systolic fabrics \cite{kim2019gpt_inference}. Profiling LLaMA3.2-1B on Jetson AGX Orin (Fig.~\ref{profiling}) shows that for input lengths of 64 to 1024 and output sequences up to 512 tokens, an average of 96.6\% of inference time is spent in decoding (batch size = 1). This motivates the need for addressing the challenges of the GEMV-heavy decoding phase, which dominates inference time.

\subsection{Compute-in-Memory}
\label{sec:cim}
CIM technology has emerged as a promising solution for efficient GEMV operations in parallel, addressing both computational and memory bottlenecks in SLM inference, particularly in the GEMV-heavy decoding phase. CIM can be realized using various memory technologies, including ReRAM \cite{hung20228,xue202116}, MRAM \cite{chiu202222nm, liu2023mdcim}, or SRAM \cite{bazzi2024reconfigurable,jssc22, bazzi2025reconfigurable, bazzi2026synthesizable}, {with SRAM-based macros standing out for their low access latency, high parallelism, low write energy, and high write endurance \cite{haensch2023compute,gul2023finfet}. 
These characteristics make SRAM-based CIM particularly suitable for decoder-only SLM inference, where repeated weight accesses and frequent data updates during autoregressive decoding favor efficient and reliable memory operations.} SRAM-based macros are typically classified into analog and digital types. Digital CIM (DCIM) avoids the analog–digital conversions and non-idealities, such as limited noise margins and process variations, that affect analog designs, thereby providing higher precision and preventing the accuracy degradation common in analog computations. DCIM generally adopts a bit-serial input approach to maximize hardware reuse and minimize area overhead, where one input bit is processed per cycle and partial results are accumulated across cycles. This approach simplifies circuit design and enables precision reconfigurability, where higher precision is achieved by increasing the number of input cycles and combining outputs across columns with shift-and-add logic. In this work, we employ a bit-serial SRAM-based DCIM macro that supports both INT4 and INT8 precisions \cite{chih202116}.

% {To complement the inference performance evaluation presented later in this paper, Table~\ref{tab:slm_quantization_accuracy} summarizes published model quality
% results for a representative subset of the SLMs considered in this work under
% Activation-aware Weight Quantization (AWQ)~\cite{lin2024awq}. In these
% evaluations, for LLaMA-3.2-1B, Qwen2.5-0.5B/1.5B,
% SmolLM2-1.7B, and Phi-3.5-mini-3.8B, zero-shot accuracy is averaged across
% BoolQ, RTE, HellaSwag, WinoGrande, ARC-Easy, ARC-Challenge, and
% OpenBookQA~\cite{zhou2025revisiting}. For Qwen3-0.6B/1.7B/4B, accuracy is
% averaged across PIQA, ARC-Easy, ARC-Challenge, HellaSwag, WinoGrande, and
% BoolQ using a separate but internally consistent evaluation
% protocol~\cite{zheng2026empirical}. Across the first model group,
% INT8 changes the average accuracy by approximately $-2.13$ to $+0.22$
% percentage points relative to the FP baseline, whereas INT4
% changes it by approximately $-4.61$ to $-0.22$ percentage points. For the
% Qwen3 models, the corresponding changes are approximately $-1.60$ to
% $+0.20$ percentage points for INT8 and $-4.60$ to $-1.30$ percentage points
% for INT4. These results indicate that both INT8 and INT4 retain accuracy close to their FP baselines, supporting their practical use for efficient SLM inference. INT8 has the smallest and most consistent changes, while INT4 remains competitive despite a modest and more model-dependent accuracy reduction, particularly for the smallest models.}

\subsection{{Quantization for Small Language Models}}
\label{sec:quantization}
{Quantization is widely used to reduce the memory footprint and computational cost of SLM inference, making it particularly important for deployment on resource-constrained edge devices. By representing model parameters with lower numerical precision, quantization reduces memory requirements and data movement, although it may introduce accuracy degradation. In this work, we consider INT4 and INT8 precisions and evaluate their implications for hardware performance. Since our focus is on the hardware implications of quantization rather than the quantization process itself, we assume that model weights have already been quantized to the target precision using an external quantization method. To contextualize the INT4 and INT8 precisions considered in this work, Table~\ref{tab:slm_quantization_accuracy} summarizes previously published model-quality results for a representative subset of the evaluated SLM families under Activation-aware Weight Quantization (AWQ)~\cite{lin2024awq}. We note that AWQ is used here as a representative example because comparable published quality results are available for several of the SLMs considered in this work; EdgeCIM itself is agnostic to the specific quantization algorithm. }

{For LLaMA3.2-1B, Qwen2.5-0.5B/1.5B, SmolLM2-1.7B, and Phi-3.5-mini-3.8B, the reported zero-shot accuracy is averaged across BoolQ, RTE, HellaSwag, WinoGrande, ARC-Easy, ARC-Challenge, and OpenBookQA~\cite{zhou2025revisiting}. For Qwen3-0.6B/1.7B/4B, accuracy is averaged across PIQA, ARC-Easy, ARC-Challenge, HellaSwag, WinoGrande, and BoolQ using the evaluation protocol in~\cite{zheng2026empirical}. The published results show smaller and generally more consistent changes for INT8, while INT4 exhibits a larger and more model-dependent quality impact. Because the entries are drawn from independent studies, differences in checkpoints, calibration procedures, and evaluation protocols should be interpreted within the context of their respective sources and not as a controlled cross-model accuracy comparison. A systematic evaluation of quantization algorithms and model-quality tradeoffs across the full SLM suite is outside the scope of this hardware architecture study.}

\begin{table}[t]
\centering
\caption{{Representative published model-quality results for selected SLMs under AWQ weight-only quantization.} INT8 and INT4 denote W8A16 and W4A16, respectively. Accuracy is averaged over the zero-shot tasks listed for each evaluation group. $\Delta$ denotes the absolute change from the floating-point baseline.}
\label{tab:slm_quantization_accuracy}
\resizebox{\columnwidth}{!}{%
\begin{tabular}{lcccccc}
\toprule
\multirow{2}{*}{Model} &
\multicolumn{2}{c}{Floating point} &
\multicolumn{2}{c}{AWQ INT8} &
\multicolumn{2}{c}{AWQ INT4} \\
\cmidrule(lr){2-3}\cmidrule(lr){4-5}\cmidrule(lr){6-7}
& Avg. (\%) & Tasks
& Avg. (\%) & $\Delta$
& Avg. (\%) & $\Delta$ \\
\midrule
\multicolumn{7}{l}{\textit{*BoolQ, RTE, HellaSwag, WinoGrande,
ARC-Easy, ARC-Challenge, and OpenBookQA}}\\
LLaMA3.2-1B       & 45.61 & 7 & 45.21 & $-0.40$ & 45.28 & $-0.33$ \\
Qwen2.5-0.5B       & 49.65 & 7 & 47.52 & $-2.13$ & 45.04 & $-4.61$ \\
Qwen2.5-1.5B       & 56.99 & 7 & 57.21 & $+0.22$ & 55.56 & $-1.43$ \\
SmolLM2-1.7B       & 56.29 & 7 & 56.29 & $\phantom{+}0.00$ & 56.07 & $-0.22$ \\
Phi-3.5-mini-3.8B  & 64.58 & 7 & 64.60 & $+0.02$ & 63.46 & $-1.12$ \\
\midrule
\multicolumn{7}{l}{\textit{**PIQA, ARC-Easy, ARC-Challenge,
HellaSwag, WinoGrande, and BoolQ}}\\
Qwen3-0.6B         & 56.50 & 6 & 56.40 & $-0.10$ & 54.80 & $-1.70$ \\
Qwen3-1.7B         & 65.60 & 6 & 64.00 & $-1.60$ & 61.00 & $-4.60$ \\
Qwen3-4B           & 68.70 & 6 & 68.90 & $+0.20$ & 67.40 & $-1.30$ \\
\bottomrule
\end{tabular}%
}
\begin{minipage}{\columnwidth}
\footnotesize
\vspace{1mm} * Results in the first group are derived from the task-level results
reported in~\cite{zhou2025revisiting}. 
** Results in the second group
are derived from~\cite{zheng2026empirical}. Small positive
differences are within normal evaluation variation and should not be
interpreted as an inherent benefit of quantization.
\end{minipage}
\end{table}

\section{Proposed Methodology}
\label{sec:methodology}
\subsection{Framework Overview}
\label{section:framework}
\begin{figure}[t]
\centering
\includegraphics[width=0.85\columnwidth]{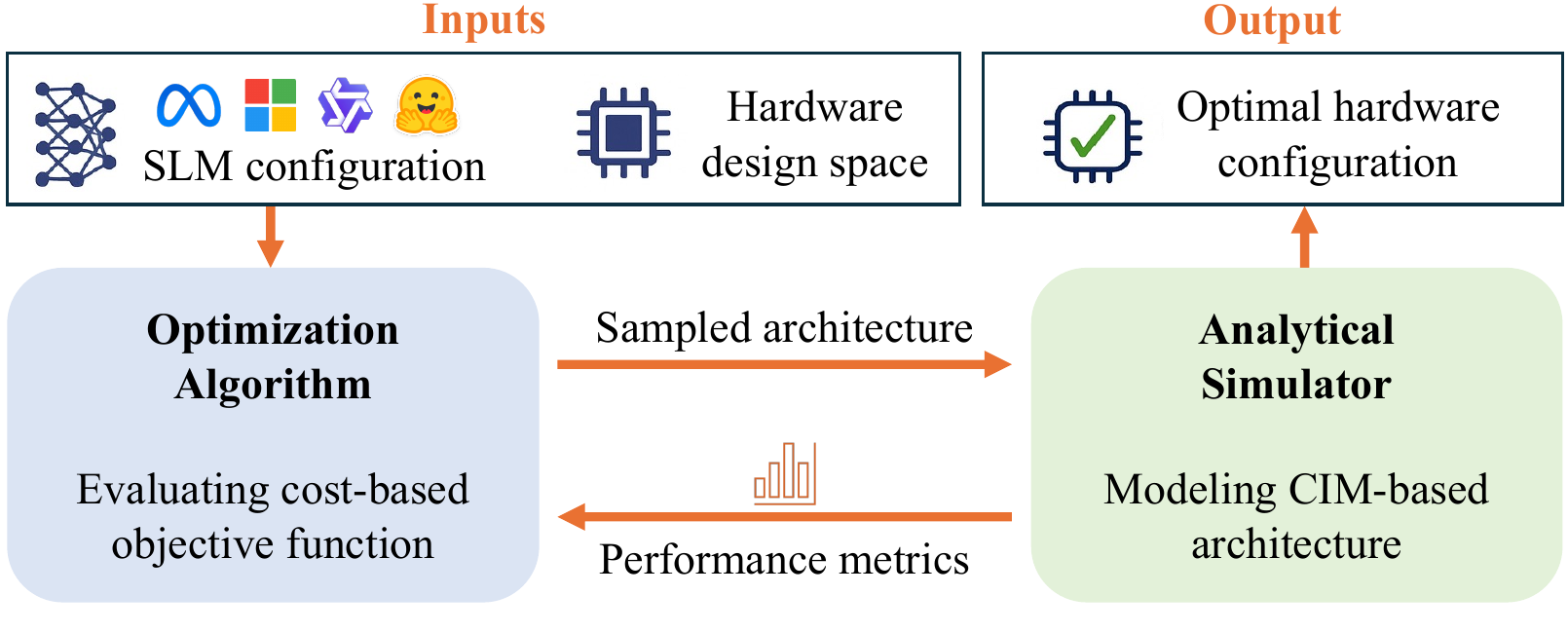}
\caption{Proposed EdgeCIM.}
\label{fig:framework}
\end{figure}

The proposed EdgeCIM is illustrated in Fig. \ref{fig:framework}. Its main objective is to explore the hardware design space and identify the optimal CIM-based hardware configuration for accelerating the decoding phase of decoder-only SLMs. The framework consists of two main components: an optimization algorithm and an analytical modeling-based simulator. It takes as input the target SLM configuration and a predefined hardware parameter search space. This search space defines the key hardware parameters considered for optimization during the DSE process. The optimization algorithm samples candidate architectures from this space and evaluates them using an objective function. Each candidate is analyzed by the simulator, which models the CIM-based architecture and reports key performance metrics. These metrics are then fed back into the optimization engine, which iteratively refines the search toward optimal solutions. The final output is the hardware configuration that minimizes the objective function, along with its optimized parameters and performance metrics. Details of the design space and the objective cost function are provided in Section \ref{sec:implementation}.

\begin{figure}[!b]
\centering
\includegraphics[width=\columnwidth]{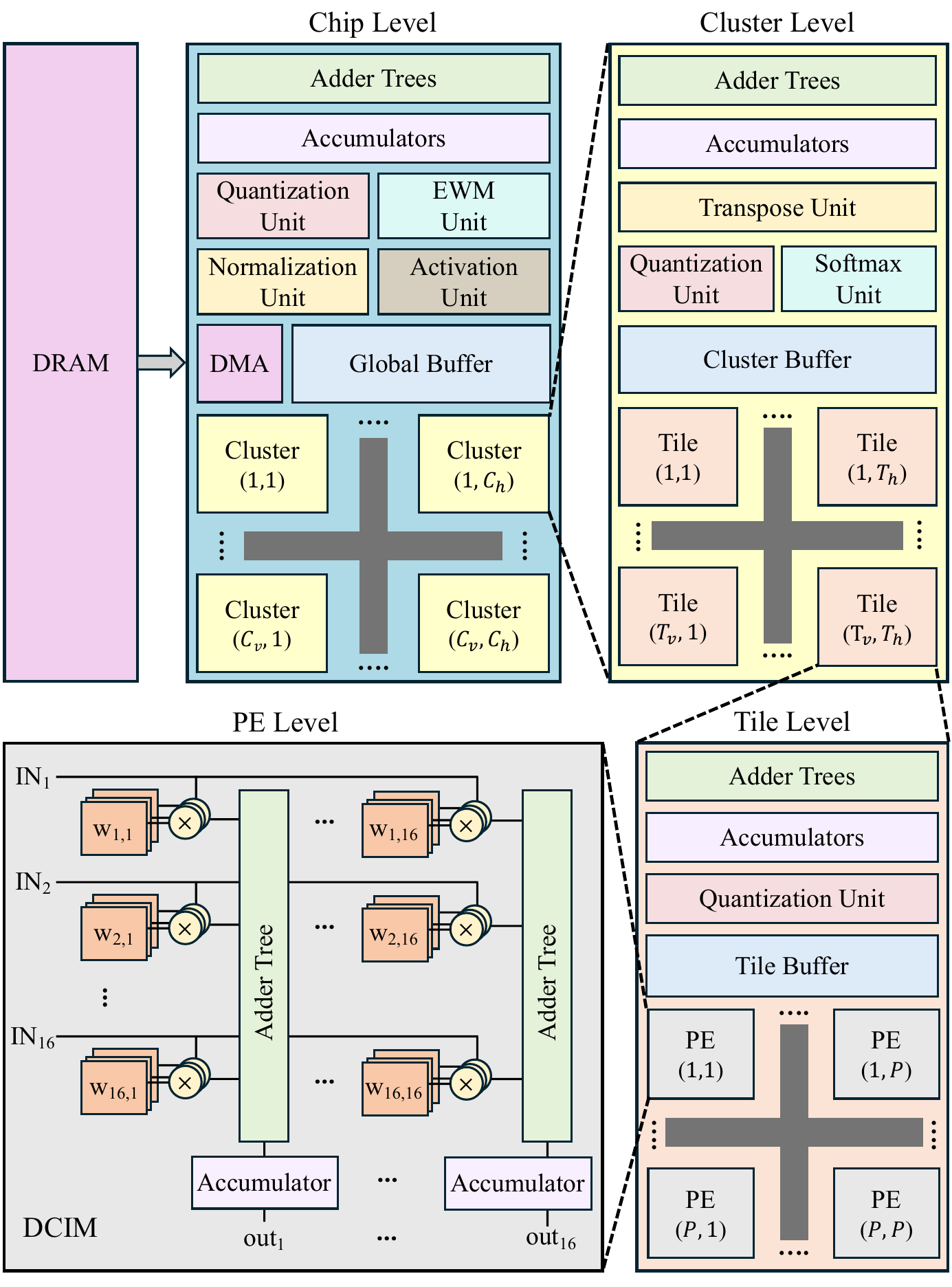}
\caption{High-level architecture of the proposed design.}
\label{fig:full_architecture}
\end{figure}

\subsection{Hardware Architecture}
The high-level architecture of EdgeCIM is shown in Fig. \ref{fig:full_architecture}. {To improve throughput, state-of-the-art CIM accelerators typically organize their compute arrays in a hierarchical manner \cite{shafiee2016isaac, ankit2019puma, kim2025hastily}. Following this, EdgeCIM adopts a tiled hierarchical architecture:} the chip consists of ($C_v\times C_h$) clusters, each cluster contains ($T_v\times T_h$) tiles, and each tile includes a ($P\times P$) array of Processing Elements (PEs). {Each PE is a 16$\times$16 SRAM-based bit-serial DCIM macro that performs GEMV using weight-stationary storage and cycle-wise accumulation \cite{chih202116}, unlike regular PEs that rely on explicit MAC units with dedicated multipliers, adders, and intermediate registers, resulting in higher energy and area overheads.} In addition to the PEs, the architecture has adder trees and accumulators to combine partial results. Buffers are employed to store intermediate and final results, while dedicated functional units handle normalization, quantization, activation, transposition, Softmax, and element-wise multiplication. A global buffer, shared across clusters, interfaces with off-chip DRAM that stores the model weights and the KV-cache. The interconnect follows a 2D hierarchical bus structure to enable efficient data transfer across the architecture. For DSE, we vary parameters including the number of clusters, the total number of tiles, the number of active tiles, the number of PEs per tile, and the bus width at each hierarchical level.

{\textbf{Architectural scope and novelty:} The individual hardware components used by EdgeCIM, including the bit-serial SRAM-CIM macro, hierarchical tiles, reduction trees, accumulators, and buffers, are established accelerator primitives. EdgeCIM instead contributes their decode-oriented organization and control. Unlike architectures that assign fixed projection and attention engines or optimize an isolated attention/FC kernel, the same EdgeCIM tile fabric is temporally reconfigured across all decoding stages. More importantly, EdgeCIM distinguishes between physical tile capacity, $T_{\mathrm{total}}$, and instantaneous execution width (or active tiles), $T_A$. As will be described in Section~\ref{section:dataflow}, only $T_A$ tiles consume the current streamed partition, while the remaining tiles are prepared for subsequent partitions. Therefore, increasing $T_{\mathrm{total}}$ does not require all instantiated tiles to generate concurrent off-chip traffic. This decoupling allows the architecture to independently trade off compute latency, partition transfer time, and overlap efficiency, and is jointly optimized with the cluster topology, PE dimensions, and interconnect widths.}

\subsection{Dataflow Mapping}
\label{section:dataflow}
\begin{figure}[t]
\centering
\includegraphics[width=\columnwidth]{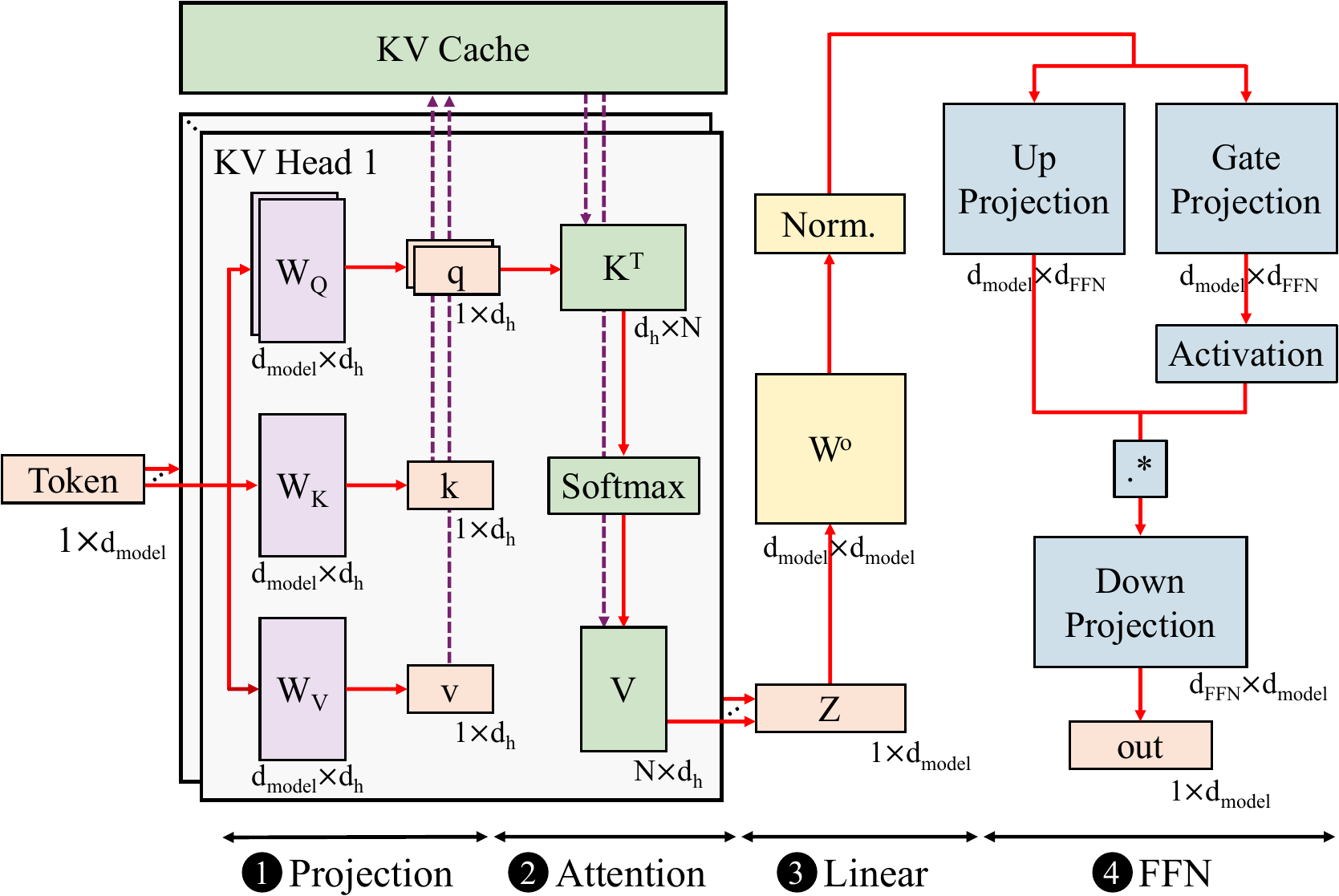}
\caption{Typical decoding phase in a decoder-only SLM.}
\label{fig:dataflow}
\end{figure}

Fig. \ref{fig:dataflow} illustrates the dataflow of the decoder-only SLM decoding phase, where the model processes one token at a time. {In this work, we focus on latency-critical, batch-size-one edge inference, where requests are processed immediately rather than accumulated for batched execution.} The decoding phase in SLMs is typically memory-bound and exhibits very low arithmetic intensity compared to other workloads \cite{kim2023full}. {The main memory bottlenecks arise from repeated
model-weight streaming during GEMV operations and
KV-cache accesses during attention computation, with
the latter becoming increasingly significant as the
context length grows.}
Since only one token is generated at a time, the
relevant KV-cache entries must be accessed for each
generated token. Moreover, with batch size one,
data reuse is minimal, meaning each KV-cache load
is used only once. As a result, memory transfers
from DRAM cannot be fully hidden by computation.
To mitigate {KV-cache access overhead},
FlashAttention~\cite{dao2022flashattention}
partitions the KV-cache into smaller blocks and
retrieves them sequentially during attention
computation. Inspired by this approach, our design
also employs a block-based strategy for the
attention mechanism. However, because we target {end-to-end decoding acceleration}, we extend this concept beyond attention to other layers such as projection, linear, and feed-forward networks. Given the limited on-chip storage and the large size of weight matrices in these layers, we partition the matrices into smaller blocks and fold computation over time. As the architecture scales with the DSE parameters, larger configurations require larger weight partitions to fully utilize the hardware. This, however, shifts the bottleneck back to off-chip DRAM transfers, as moving large partitions dominates runtime and reduces the overlap of data transfer with computation. To alleviate this imbalance and improve pipeline efficiency, we introduce the notion of active tiles, where only a subset of the total tiles, denoted as $T_A$, is used in parallel. At any given time, data is transferred only for these $T_A$ tiles, reducing the DRAM transfer size per pipeline stage. While computation proceeds on the active tiles, weights for the remaining tiles are prefetched, effectively overlapping transfer and compute. {
Specifically, for a streamed weight partition requiring
$N_W(T_A)$ quantized weights, the communication and computation times of
one pipeline interval are approximated as
\begin{equation}
t_{\mathrm{mem}}(T_A)
=
\frac{N_W(T_A)\,p_W}
{BW_{\mathrm{DRAM}}},
\end{equation}
and
\begin{equation}
t_{\mathrm{comp}}(T_A)
=
\frac{N_{\mathrm{op}}(T_A)}
{\Theta_{\mathrm{CIM}}(T_A)},
\end{equation}
where $p_W$ is the weight precision,
$BW_{\mathrm{DRAM}}$ is the available off-chip DRAM bandwidth,
$N_{\mathrm{op}}(T_A)$ is the number of operations performed by the
active tiles for the current partition, and
$\Theta_{\mathrm{CIM}}(T_A)$ is the aggregate throughput of the active
CIM tiles. When all tiles are active
($T_A=T_{\mathrm{total}}$), no inactive tiles remain to preload the next
weight partition, and communication and computation proceed
sequentially. In contrast, under the proposed active-tile scheduling
($T_A<T_{\mathrm{total}}$), the inactive tiles preload the next weight
partition while the active tiles execute the current partition, yielding
a steady-state pipeline interval of
\begin{equation}
t_{\mathrm{stage}}(T_A)
\approx
\max\!\left(
t_{\mathrm{mem}}(T_A),
t_{\mathrm{comp}}(T_A)
\right).
\end{equation}
Increasing the number of active tiles improves computational throughput
by increasing $\Theta_{\mathrm{CIM}}(T_A)$, but also increases the
amount of weights that must be transferred during each pipeline
interval. EdgeCIM therefore treats $T_A$ as a design
parameter and uses the DSE framework to identify the value that best
balances communication and computation for each model,
precision, and memory-bandwidth constraint.
}

Since the SLM decoding process consists of sequential stages, where the
output of each stage becomes the input to the next, as shown in
Fig.~\ref{fig:dataflow}, these stages are mapped sequentially onto the
proposed hardware, as described below.

\subsubsection{Projection} 
\begin{figure}[!b]
\centering
\includegraphics[width=\columnwidth]{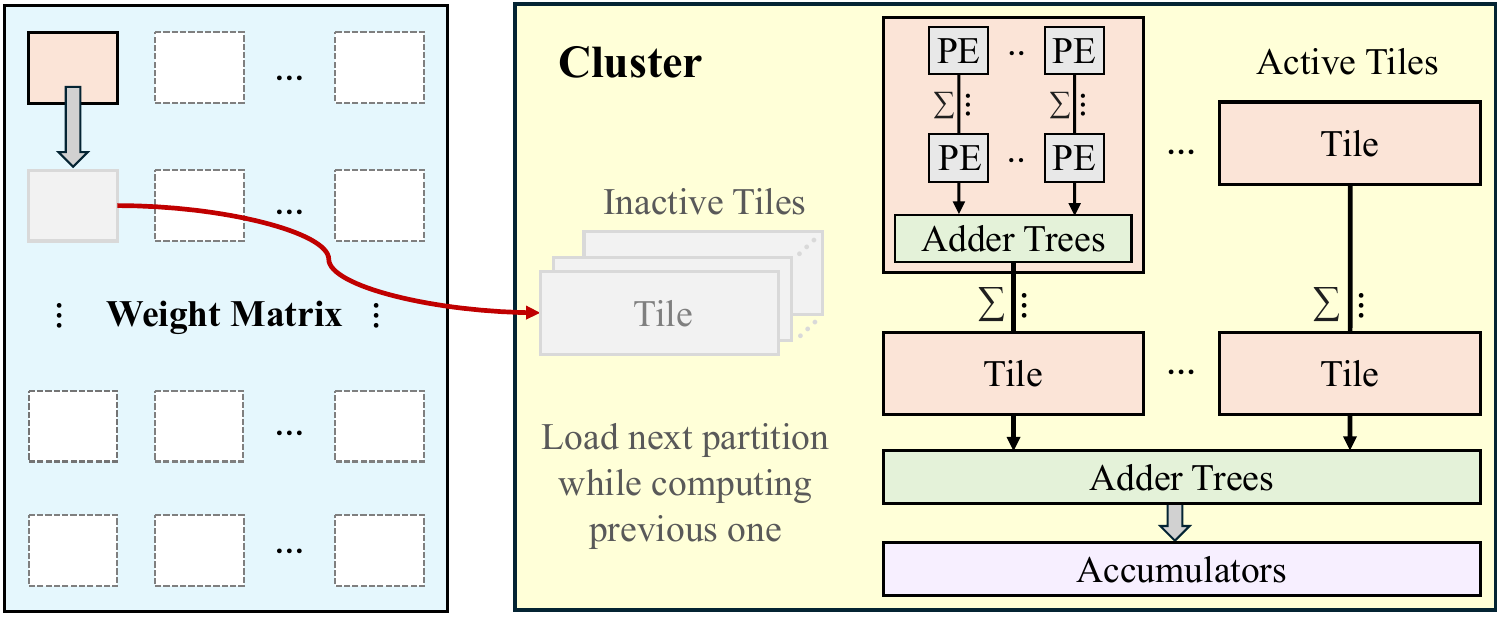}
\caption{Partition-based mapping of weights onto EdgeCIM.}
\label{fig:partitions}
\end{figure}

In this phase, the query ($q$), key ($k$), and value ($v$) vectors are generated by multiplying the input token with the projection matrices $W_Q$, $W_K$, and $W_V$, respectively. In our mapping, each head (or KV head in GQA) is assigned to a cluster, with all clusters operating in parallel. If the number of heads exceeds the number of available clusters, they are processed sequentially. Within each cluster, the projection matrices $W_{Q,K,V}$ are also processed sequentially. Each matrix is divided into partitions that are streamed from DRAM one at a time and mapped onto the tiles of the assigned cluster, while the input token is broadcast to all clusters. Computation and data transfer are overlapped using the active-tile mechanism introduced earlier: active tiles process the current partition while others preload the next one, as illustrated in Fig. \ref{fig:partitions}. Within each tile, the PEs perform GEMV operations between the assigned weight partitions and the corresponding portion of the input token vector. Partial sums from vertically aligned PEs are reduced by tile-level adder trees, and the resulting outputs are further aggregated vertically across tiles at the cluster level. Results from vertically partitioned submatrices of the same weight matrix are accumulated using accumulators, and outputs from horizontally partitioned submatrices are concatenated to form the complete result. After computation, the key and value vectors are quantized, transposed, and written back to DRAM to be appended to the KV-cache, while the query results are quantized and stored in the cluster buffer to serve as inputs in the subsequent phase.

\subsubsection{Attention} Following the projection phase, the attention mechanism computes the output $\text{Softmax}(qK^T)V$. Each attention head is mapped to a cluster, with keys and values streamed from DRAM in block partitions of size $(b \times d_h)$. Queries are broadcast from the cluster buffer to all tiles. In the case of GQA, grouped queries reuse the same keys and values within each KV head in a pipelined fashion before fetching the next block from DRAM. Within each cluster, tiles are divided between key and value processing. PEs first perform GEMV operations between the query and the stored key block. Tile- and cluster-level adder trees then aggregate these GEMV outputs to produce block-level attention scores $qK^T_{\text{block}}$, which are processed by a dedicated Softmax unit operating in a block-wise manner, following \cite{dao2022flashattention}. The resulting attention scores are multiplied with $V_{\text{block}}$, added across the block dimension $b$ using tile- and cluster-level adder trees, and accumulated across loaded blocks using accumulators. The final attention output from each head is then quantized and written to the global buffer, where it is concatenated with the outputs of the other heads.

\subsubsection{Linear} The output projection matrix $W^o$ is partitioned and mapped across the PEs of all clusters in parallel, in contrast to the projection and attention phases where clusters operate independently. Chip-level adder trees and accumulators are employed to aggregate partial results across clusters and to accumulate outputs across partitions. After multiplication, the outputs are normalized using a dedicated hardware unit and stored in the global buffer to be used in the next stage.

\subsubsection{Feed-Forward Network} Finally, the feed-forward network maps the up and gate projection matrices in partitions across clusters, with PEs divided between the two in parallel. The outputs of the gate projection are passed through a dedicated activation unit and then multiplied element-wise with the corresponding outputs of the up projection. Once all partitions of the up and gate matrices have been mapped and combined, the resulting vector is processed by the down projection matrix, which is likewise partitioned and mapped in sequence. The final output is stored in the global buffer.

\textbf{Auxiliary Operators:} Activation, quantization, transposition, normalization, and Softmax are executed on dedicated hardware units within our architecture, and their performance overhead is incorporated into the analysis.

\section{Hardware--Software Co-Optimization Process}
\label{sec:implementation}

\subsection{Design Space Exploration Framework}
\label{sec:dse}
We now formalize the optimization problem and describe the hardware--software co-optimization process in EdgeCIM. Given a predefined hardware design space $\mathcal{H}$, the objective is to determine the optimal CIM-based hardware configuration $h^*\in \mathcal{H}$ that minimizes a cost function capturing the trade-off between latency and energy for executing the decoding phase of a decoder-only SLM. The problem is expressed as:
\begin{equation}
\underset{h \in \mathcal{H}}{\text{minimize}} \; 
L(h)^{\alpha} \times E(h)^{(1-\alpha)}, 
\quad 0 \leq \alpha \leq 1
\end{equation}
{where $L(h)$ and $E(h)$ denote the latency and energy of generating a specified number of tokens during autoregressive decoding under configuration $h$, assuming a specified number of prefill tokens as the initial context.} The parameter $\alpha$ is tunable and explores the latency-energy trade-off, enabling the framework to prioritize either latency or energy depending on deployment requirements. {No explicit normalization of latency and energy is required, as changing their units scales the objective by a constant factor without affecting the relative ranking of hardware configurations.} {The cost function considers only latency and energy, while area is evaluated separately. A reference area budget of 105~mm\textsuperscript{2} is adopted, and all configurations selected through DSE satisfy this budget, as demonstrated in Section~\ref{sec:evaluation}.} We solve this problem using a genetic algorithm (GA) implemented in Python with 50 generations and a population size of 20. The GA initializes with random configurations sampled from $\mathcal{H}$, which includes the number of vertical and horizontal clusters: $C_v,C_h\!\in\!\{1,\ldots,5\}$, active tiles per cluster: $T_A=T_v^{\text{act}}\times T_h^{\text{act}}$ with $T_v^{\text{act}},T_h^{\text{act}}\!\in\!\{2,\ldots,8\}$, total tiles per cluster: $T_{\text{total}}=T_v\times T_h=M\times T_A$ with $M\!\in\!\{1,\ldots,8\}$, processing elements per tile: $P^2\!\in\!\{4,9,16,25,36\}$, on-chip bus widths, including inter-cluster, inter-tile, and intra-tile buses $\in \{512, 1024, 2048, 4096\}$. Overall, this search space contains $\sim3.1\times10^6$ configurations. Each configuration is then evaluated by an analytical simulator implemented in C++ with hardware components modeled as dedicated classes. The simulator models the hierarchical organization of the architecture (PE, tile, cluster, and chip levels) and evaluates the execution of the mapped dataflow for each candidate hardware configuration. {For each mapped layer, the simulator analytically derives the required computation and data-movement operations from the matrix partitioning and the selected hardware configuration. The execution model explicitly captures CIM array loading, bit-serial computation, partial-sum accumulation, buffer accesses, interconnect transfers, and off-chip memory transactions. The latency and energy of each operation are obtained from the corresponding component-level models and accumulated according to the modeled pipeline dependencies to estimate the end-to-end execution latency and energy, while also reporting the corresponding hardware area.} Configurations are ranked by the cost function, and new candidates are generated by simulated binary crossover and polynomial mutation (crossover probability = 1, distribution index = 3). This process iterates over generations, converging toward better solutions, and the best configuration $h^*$ is selected as the optimal architecture.
% \textcolor{red} {A DSE run exploring 1,000 hardware configurations per SLM required approximately 33 seconds on average across the evaluated models, using eight parallel workers on an Apple M3 (8 CPU cores, 24 GiB RAM), including initialization, compilation, and optimization.}

For component-level modeling, we use CACTI 6.0 (65~nm) \cite{cacti} to estimate the latency, energy, and area of buffers, while compute components are characterized using HSPICE \cite{hspice} at the same technology node. The interconnect parameters are adopted from \cite{mesh}. {Off-chip memory is modeled as LPDDR5X DRAM with 16 channels, corresponding to a peak theoretical bandwidth of approximately 273~GB/s. This represents an upper-end configuration close to commercially used edge-memory configurations: NVIDIA Jetson AGX Orin provides up to 204.8~GB/s through a 256-bit LPDDR5 interface~\cite{nvidia_jetson_thor_datasheet,nvidia_jetson_orin}. Laptop-class edge platforms also employ high-bandwidth LPDDR5X; for example, Qualcomm Snapdragon X Elite supports up to 135~GB/s~\cite{qualcomm_snapdragon_x_elite}. }

\begin{figure}[!b]
    \centering
    \subfloat[]{\includegraphics[width=0.95\columnwidth]{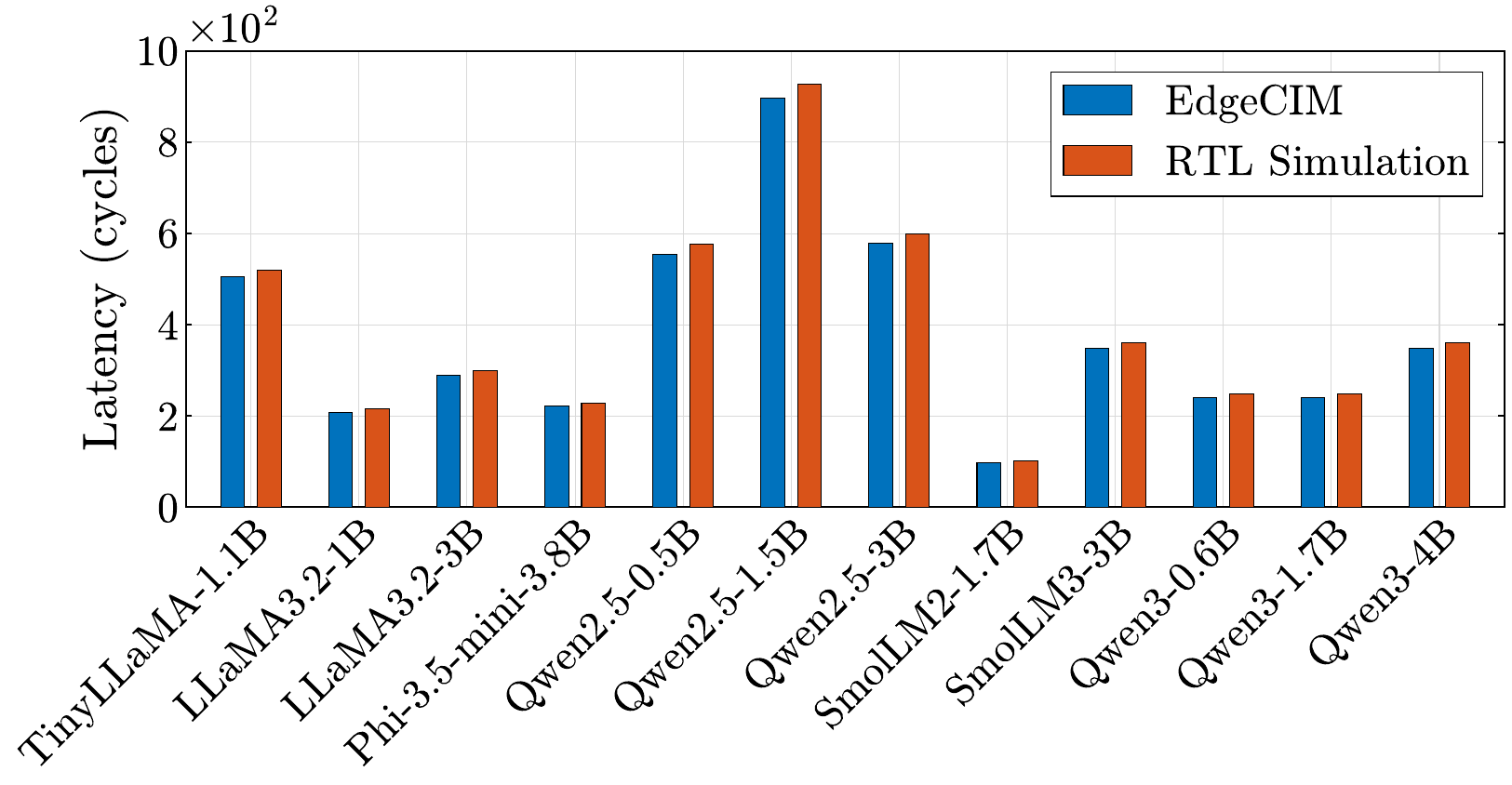}}\hfill
    \subfloat[]{\includegraphics[width=0.95\columnwidth]{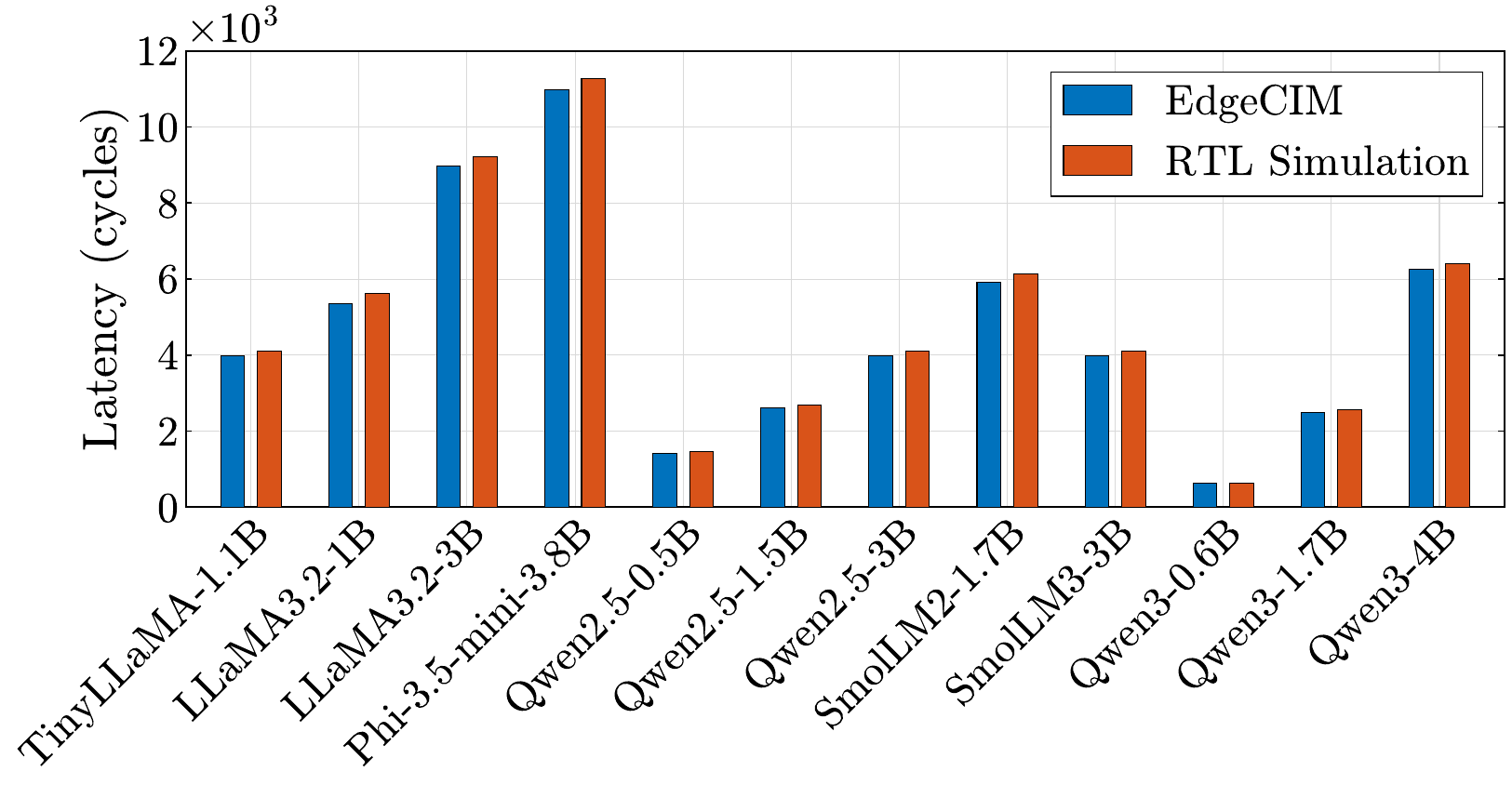}}\hfill

    \caption{{Performance model validation against RTL simulations for
(a) attention and (b) output-projection kernels.}}
    \label{fig:rtl}
\end{figure}

{\textbf{Simulator Validation:} Given the pre-silicon, architecture-level design-space exploration
nature of EdgeCIM, we assess the accuracy of its analytical models
through validation at both the processing-element and kernel
levels. The PE model adopts the bit-serial SRAM-CIM architecture of the fabricated TSMC implementation reported in~\cite{chih202116}. After technology scaling and configuration matching, the modeled PE exhibits differences of 7.74\% in energy efficiency and 5.26\% in area efficiency relative to the fabricated implementation. The scaling is performed using DeepScaleTool~\cite{sarangi2021deepscaletool}, which derives scaling factors from published silicon technology trends rather than ideal constant-field scaling. At the kernel level, the analytical
execution model is compared against cycle-level RTL simulations
of the output-projection and attention kernels using the
DSE-selected hardware configurations. As shown in
Fig.~\ref{fig:rtl}, the estimated execution cycles differ from
RTL by 1.7\% to 5.4\% for output projection and
2.8\% to 4.1\% for attention. These kernels exercise distinct
execution paths, covering partitioned GEMV computation,
partial-sum reduction and accumulation, hierarchical data
movement, and KV-cache accesses.}

\subsection{Design Space Exploration Analysis}
\label{dse_analysis}
\begin{figure}[t]
    \centering
    \subfloat[]{\includegraphics[width=0.45\columnwidth]{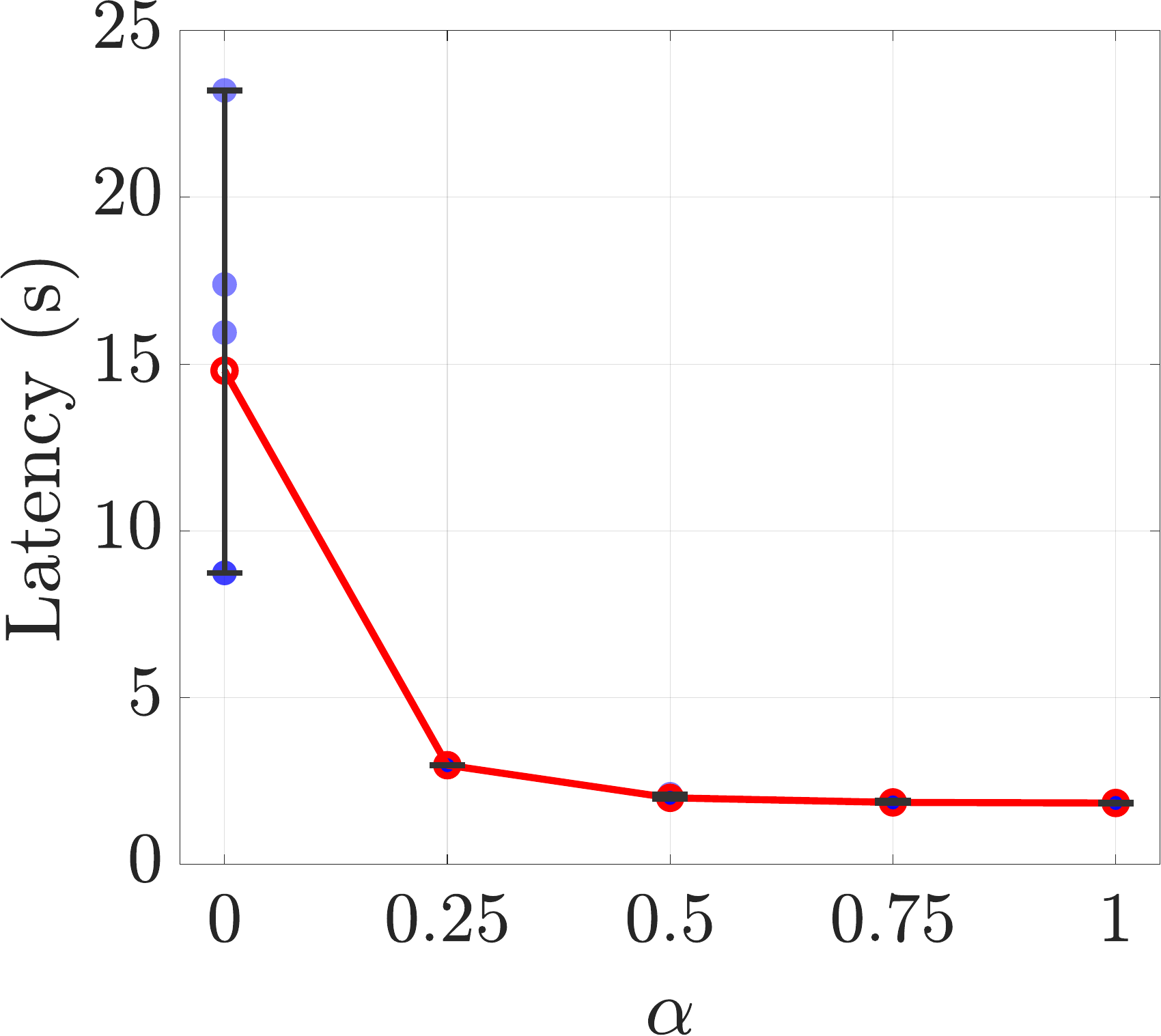}}\hfill
    \subfloat[]{\includegraphics[width=0.45\columnwidth]{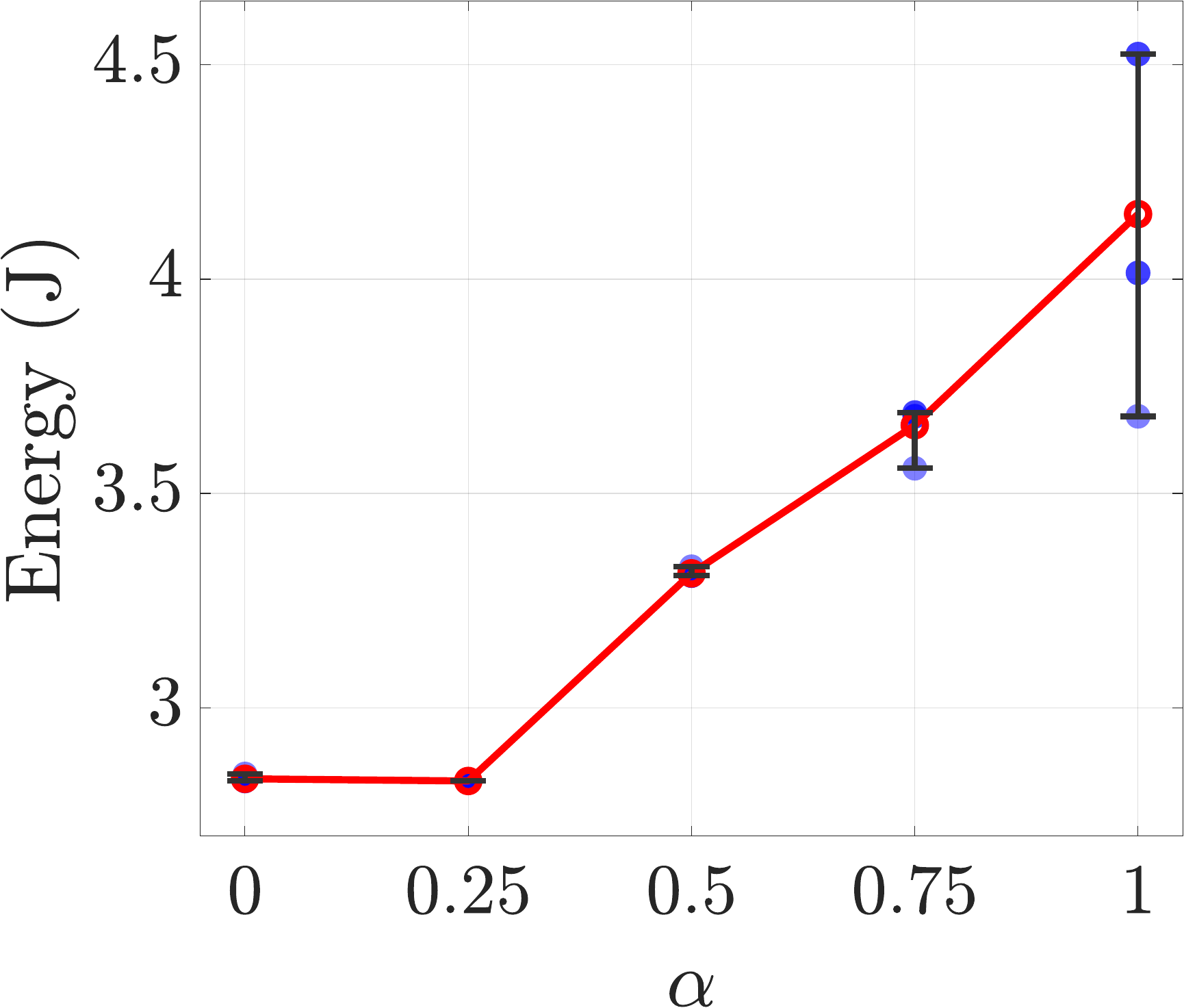}}
    \subfloat[]{\includegraphics[width=0.445\columnwidth]{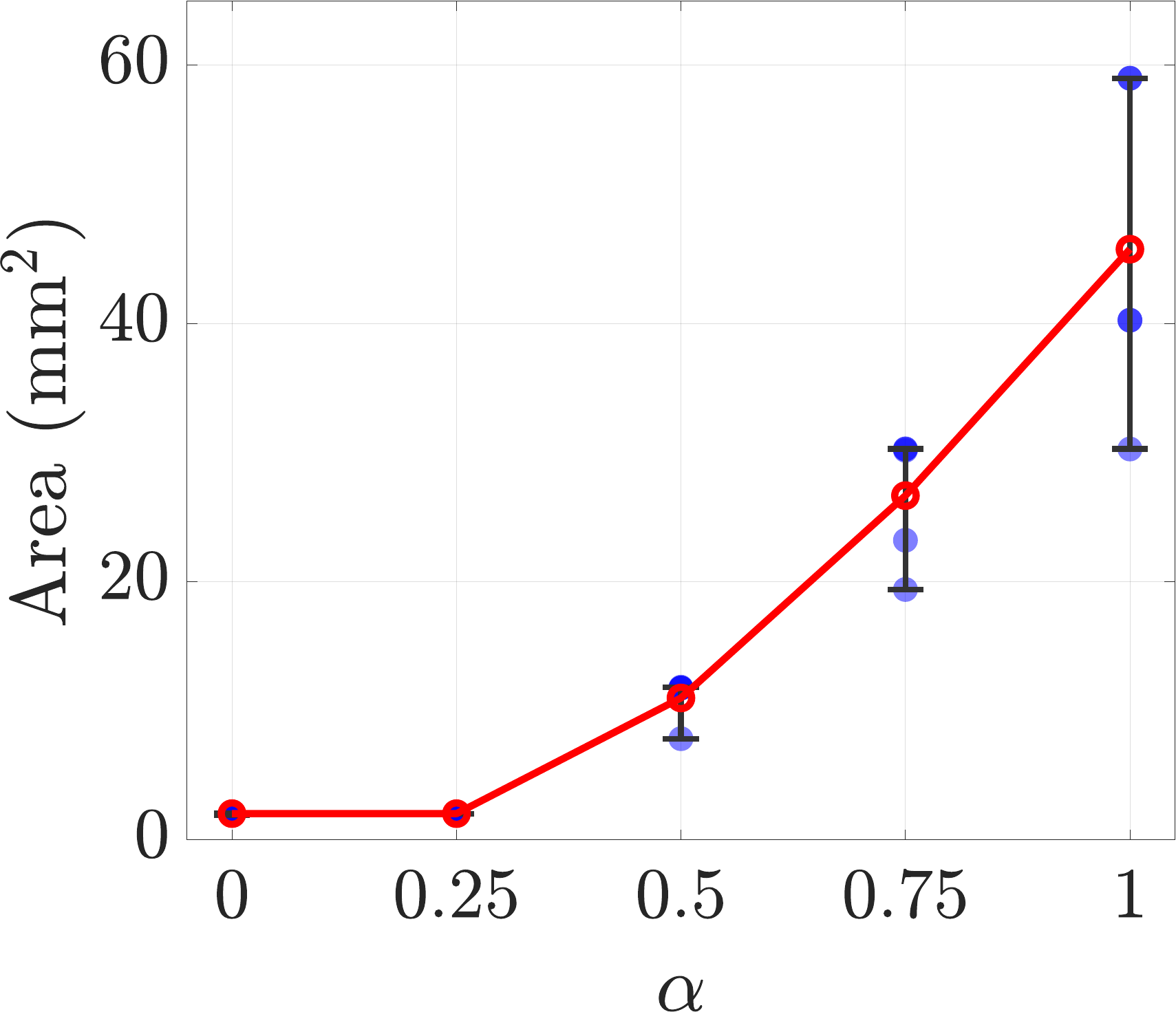}}
    \caption{{(a) Latency, (b) energy, and (c) area of the DSE-selected configurations across different values of $\alpha$ for LLaMA3.2-3B (INT8) with a 128-token prefill and 128 generated tokens. Red markers denote the average results over five independent GA runs.}}
    \label{fig:alpha_sweep}
\end{figure}

To explore the latency-energy trade-off captured by the cost function, we apply the proposed DSE framework to the LLaMA3.2-3B model with INT8 precision across different values of $\alpha$. The evaluation is performed for generating 128 tokens with 128 prefill tokens. For each $\alpha$, the framework selects the hardware configuration that minimizes the cost function at that value. The resulting latency, energy, and hardware area trends are shown in Fig.~\ref{fig:alpha_sweep}(a)--(c), respectively. To account for randomness in the search process, we perform five independent GA runs for each value of $\alpha$. The figure presents the results of all five runs, with their average highlighted in red. We observe that variability across runs differs with $\alpha$, with the most pronounced variations occurring at the extremes. At $\alpha=0$, the search is indifferent to latency, allowing designs with similar energy but widely varying latency to emerge. At $\alpha=1$, the opposite behavior is observed, where designs converge to comparable latency yet span a wide range of energy values. For intermediate $\alpha$ values, the joint penalty on both metrics narrows the feasible region, concentrating solutions near the Pareto knee and reducing variability. As $\alpha$ increases overall, the optimization progressively prioritizes latency over energy, leading to a sharp reduction in latency at the cost of higher energy consumption. Conversely, smaller $\alpha$ values emphasize energy efficiency but result in significantly higher latency. {Fig. \ref{fig:alpha_sweep}(c) reports the hardware area of the DSE-selected configurations. As $\alpha$ increases, the DSE selects increasingly parallel hardware configurations with more clusters, tiles, and processing elements to reduce execution latency, resulting in a larger hardware area. Nevertheless, all selected configurations remain within practical area budgets for edge deployment, supporting our decision to exclude area from the optimization objective.}

\begin{figure}[t]
\centering
\includegraphics[width=1\columnwidth]{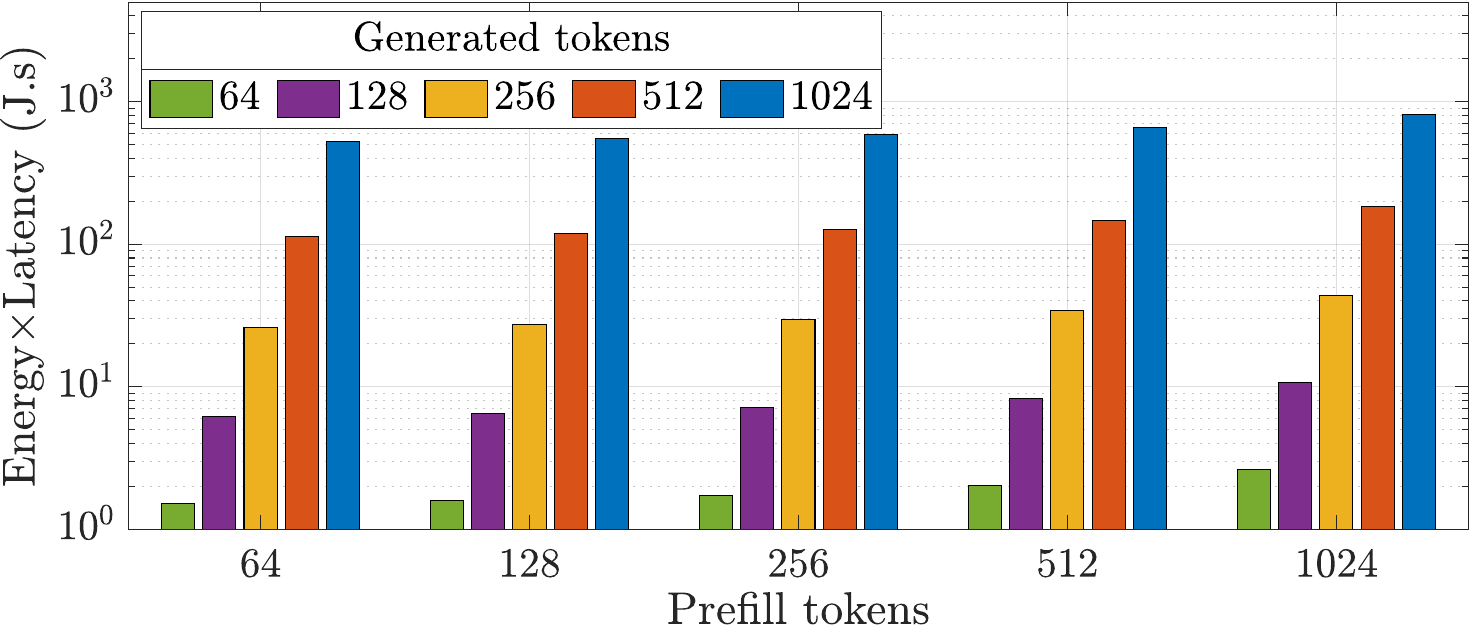}
\caption{Decoding phase energy-latency product for LLaMA3.2-3B (INT8) using the optimal $h^*$ at $\alpha = 0.5$.}
\label{fig:prefill_seqLength}
\end{figure}

To further assess the performance of the CIM-based accelerator, we investigate the impact of sequence length and prefill tokens on latency and energy. We fix $\alpha = 0.5$ to achieve a balanced trade-off between the two metrics. For LLaMA3.2-3B (INT8),
% the GA converges to the optimal configuration $h^*$: $C_v{=}2$, $C_h{=}3$, $T_v^{\text{act}}{=}4$, $T_h^{\text{act}}{=}2$, $T_{\text{total}}{=}8$, $P^2{=}4$, and bus widths $\{\text{inter-tile},\text{intra-tile},\text{inter-cluster}\}{=}\{4096,4096,4096\}$, occupying $\sim$11.83mm$^2$. Using this accelerator, 
Fig.~\ref{fig:prefill_seqLength} shows how the energy-latency product scales with prefill and generated tokens. The cost grows rapidly with the number of generated tokens, since each additional token must be sequentially processed through all layers of the model and depends on the computations of all previously generated tokens. Increasing the number of prefill tokens also raises the cost; however, the effect is less pronounced than that of decoding tokens. {This is because the reported latency and energy correspond to the autoregressive decoding phase and do not include prefill execution. The prefill tokens instead determine the initial KV-cache occupancy at the start of decoding. During decoding, their influence is limited to the attention mechanism, where a larger KV-cache must be loaded and accessed.}

% This is because prefill tokens primarily impact the prefill stage of inference, whereas our evaluation focuses on decoding. During decoding, the influence of prefill tokens is limited to the attention mechanism, where a larger KV-cache must be loaded and accessed.

\section{Results and Analysis}
\label{sec:results}
Building on the previous analysis, we fix $\alpha=1$ to prioritize latency, as EdgeCIM consistently achieves low energy across all settings. Unless otherwise stated, all experiments use a batch size of one, 128 prefill tokens, and 128 generated tokens. The benchmark suite comprises TinyLLaMA-1.1B~\cite{zhang2024tinyllama}, LLaMA3.2-1B and LLaMA3.2-3B~\cite{meta2024llama32}, Phi-3.5-mini-3.8B~\cite{abdin2024phi3technicalreporthighly}, Qwen2.5-0.5B, Qwen2.5-1.5B, and Qwen2.5-3B~\cite{qwen2025qwen25technicalreport}, SmolLM2-1.7B~\cite{allal2025smollm2}, SmolLM3-3B~\cite{smollm3blog2025}, and Qwen3-0.6B, Qwen3-1.7B, and Qwen3-4B~\cite{yang2025qwen3}. We first characterize the off-chip memory bottleneck that motivates the proposed active-tile scheduling strategy and then quantify the latency benefit of this strategy. We then evaluate EdgeCIM in terms of end-to-end throughput, energy efficiency, and hardware area across the benchmark suite, followed by comparisons with commercial edge GPUs and NPUs as well as prior CIM-based accelerators.

\subsection{{Memory-System Analysis and Active-Tile Scheduling}}
\label{sec:activeTiles}

{We first quantify the data-movement bottleneck during decoder-only SLM inference. For each model in the benchmark suite, we analyze the DRAM weight traffic under its DSE-selected hardware configuration, accounting for the effects of weight partitioning, tiling, and other implementation overheads introduced by mapping billion-parameter SLMs onto finite-capacity SRAM-CIM arrays. Across the evaluated models, the required DRAM weight traffic ranges from {357.83~MB/token to 3.63~GB/token}, corresponding to {1,680 to 20,304} DRAM weight-block fetches per generated token. These results confirm that repeated weight streaming remains a dominant communication cost during decoder-only inference.} 

{In addition to repeated weight streaming, KV-cache traffic
increases with context length as newly generated tokens
expand the cache accessed during subsequent decoding steps.
At INT8 precision, for 128 generated tokens, the average
KV-cache traffic ranges from {1.18 to 37.85~MB/token}
for a 128-token prefill to
{201.72~MB to 6.46~GB/token} for a 32K-token prefill.
Consequently, KV-cache accesses account for up to
{81.04\%} of the combined weight and KV-cache traffic
at a 32K-token prefill, highlighting their increasing
contribution to long-context decoding. {The corresponding DRAM capacity required to store
model weights and the final KV-cache ranges from
0.5 to 4.04~GB and 0.7 to 10.29~GB, respectively.}

{Considering both repeated weight streaming and KV-cache accesses, the effective DRAM bandwidth utilization ranges from 40.6\% to 91.8\%, while DRAM accesses account for 37.6\% to 69.2\% of the total inference energy across the DSE-selected configurations. These results establish that decoder-only SLM inference remains fundamentally memory-bound and motivate the proposed active-tile strategy, which reduces exposed communication latency by overlapping data transfer with computation.}

{Across the evaluated configurations, the global,
cluster, and tile buffer capacities range from
70 to 512~KiB, 7 to 64~KiB per cluster, and 0.5 to 4~KiB
per tile, respectively. During active-tile scheduling,
the current weight partition resides in the SRAM-CIM
arrays of active tiles, while the next partition is
prefetched into the inactive tiles.}

{To quantify the effectiveness of this scheduling
strategy in reducing exposed communication latency,
we perform a decoding-latency sensitivity analysis
using the DSE-selected hardware configuration for
Phi-3.5-mini-3.8B at INT8 precision. {The selected configuration contains
$T_{total}=16$ tiles per cluster, while
the DSE identifies $T_A=8$ active tiles per cluster
as the optimal operating point. To further investigate
the impact of active-tile parallelism, we perform a
post-DSE sensitivity analysis by varying
$T_A=\{2,4,6,8,12,14,16\}$, including values
outside the original DSE search space, while keeping
all other hardware parameters fixed.} The resulting decoding latency is shown in Fig.~\ref{fig:active_tiles1}. As the number of active tiles increases from 2 to 8, the decoding latency decreases from 2.804~s to 2.285~s, since additional active tiles improve computational parallelism while preserving sufficient inactive tiles to preload the next weight partition. Beyond this point, latency increases because activating more tiles reduces the number of inactive tiles available for prefetching, thereby diminishing communication--computation overlap. Consequently, the latency increases to 3.53~s, 4.31~s, and 4.436~s for $T_A=12$, 14, and 16, respectively. Overall, the DSE-selected configuration achieves a 1.94$\times$ speedup over the all-active baseline.}

\begin{figure}[b]
    \centering
    \includegraphics[width=0.7\columnwidth]{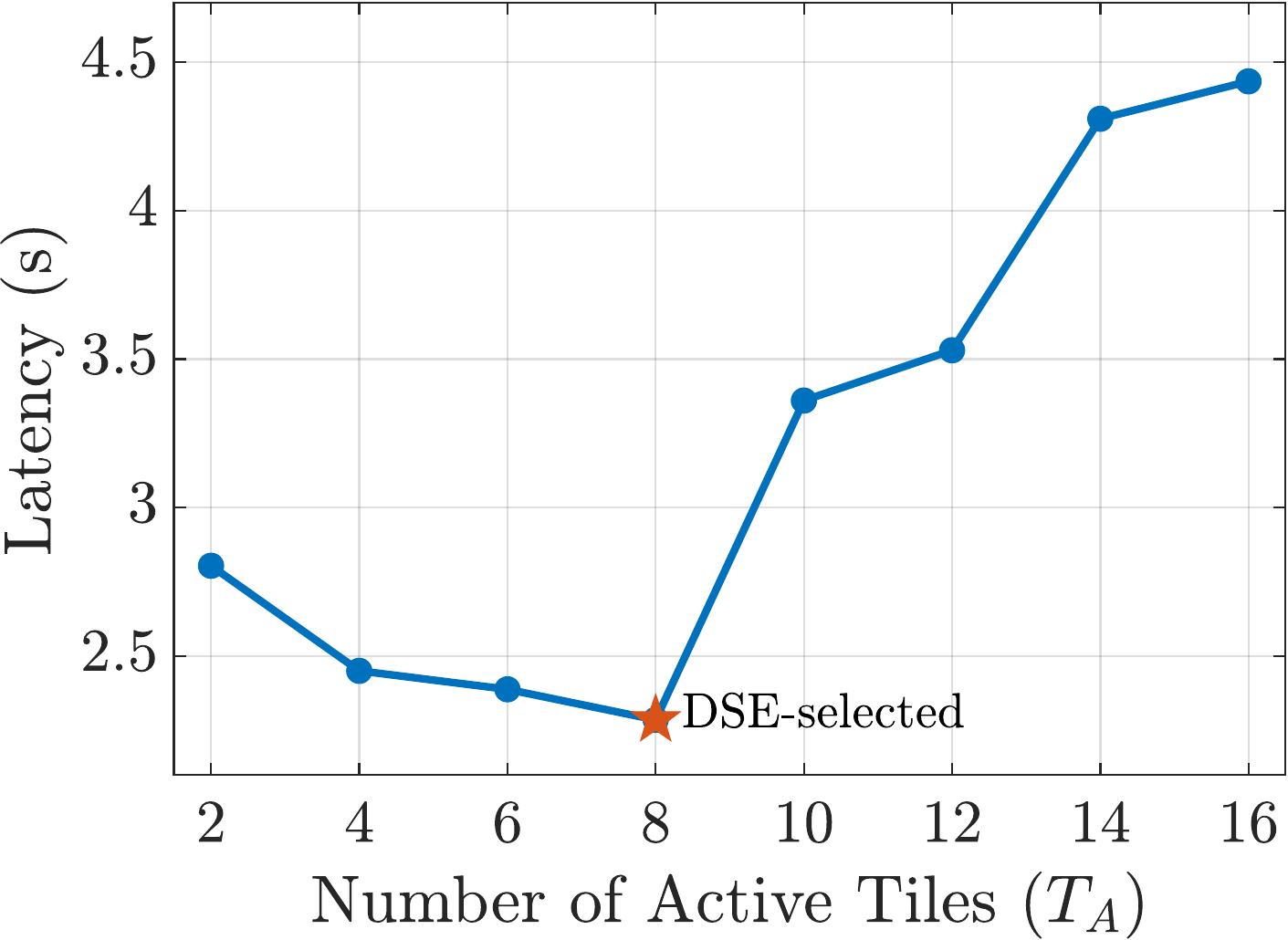}
    \caption{{End-to-end decoding latency as a function of the number of active tiles for Phi-3.5-mini-3.8B (INT8).}}
    \label{fig:active_tiles1}
\end{figure}

{To evaluate whether this behavior generalizes across different SLMs, Fig.~\ref{fig:active_tiles2} compares the DSE-selected active-tile configuration against the all-active baseline for all evaluated models at INT4 precision. Across all models, the proposed scheduling consistently achieves a 1.48$\times$ to 1.81$\times$ speedup, with an average improvement of 1.71$\times$. These results confirm that selecting only a subset of tiles for computation effectively balances computational parallelism and communication-computation overlap, thereby reducing the exposed DRAM latency during memory-bound decoder-only SLM inference.}

{We further evaluate DRAM bandwidth sensitivity while
keeping the DSE-selected hardware configurations fixed.
Relative to the nominal
273~GB/s baseline, reducing the aggregate DRAM bandwidth
to 204.8, 136.5, and 68.3~GB/s increases the decoding
latency by 1.14$\times$, 1.45$\times$, and 2.38$\times$,
respectively, for 128 prefill tokens and 128 generated
tokens. These results highlight the sensitivity of decoding performance to DRAM bandwidth, while also reflecting the contribution of on-chip processing and computation--communication overlap.}

\begin{figure}[b]
    \centering
    \includegraphics[width=0.9\columnwidth]{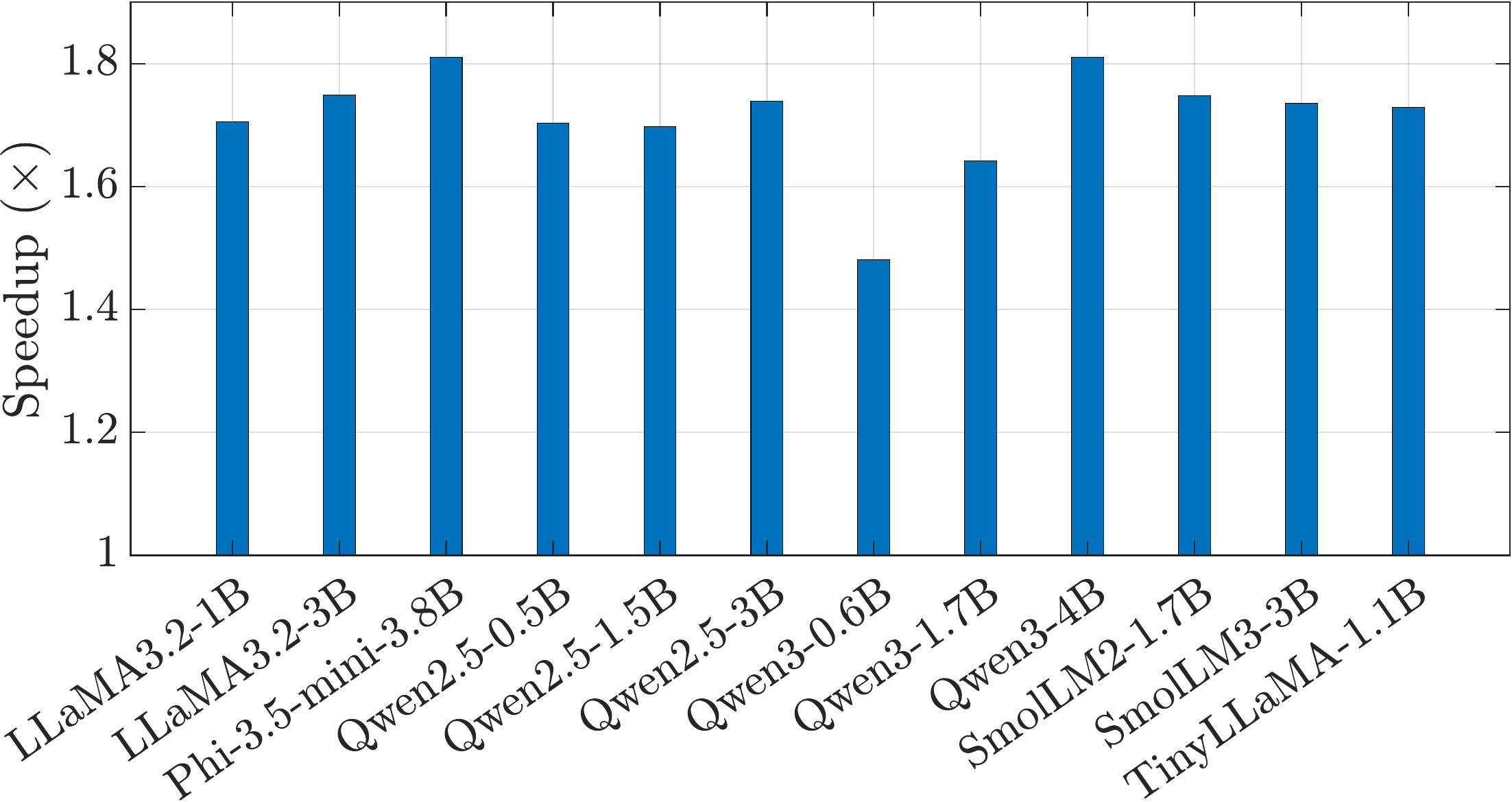}
    \caption{{End-to-end decoding speedup of the DSE-selected active-tile configuration over the all-active baseline across the evaluated SLMs (INT4).}}
    \label{fig:active_tiles2}
\end{figure}

\begin{figure}[t]
    \centering
    \subfloat[]{\includegraphics[width=0.95\columnwidth]{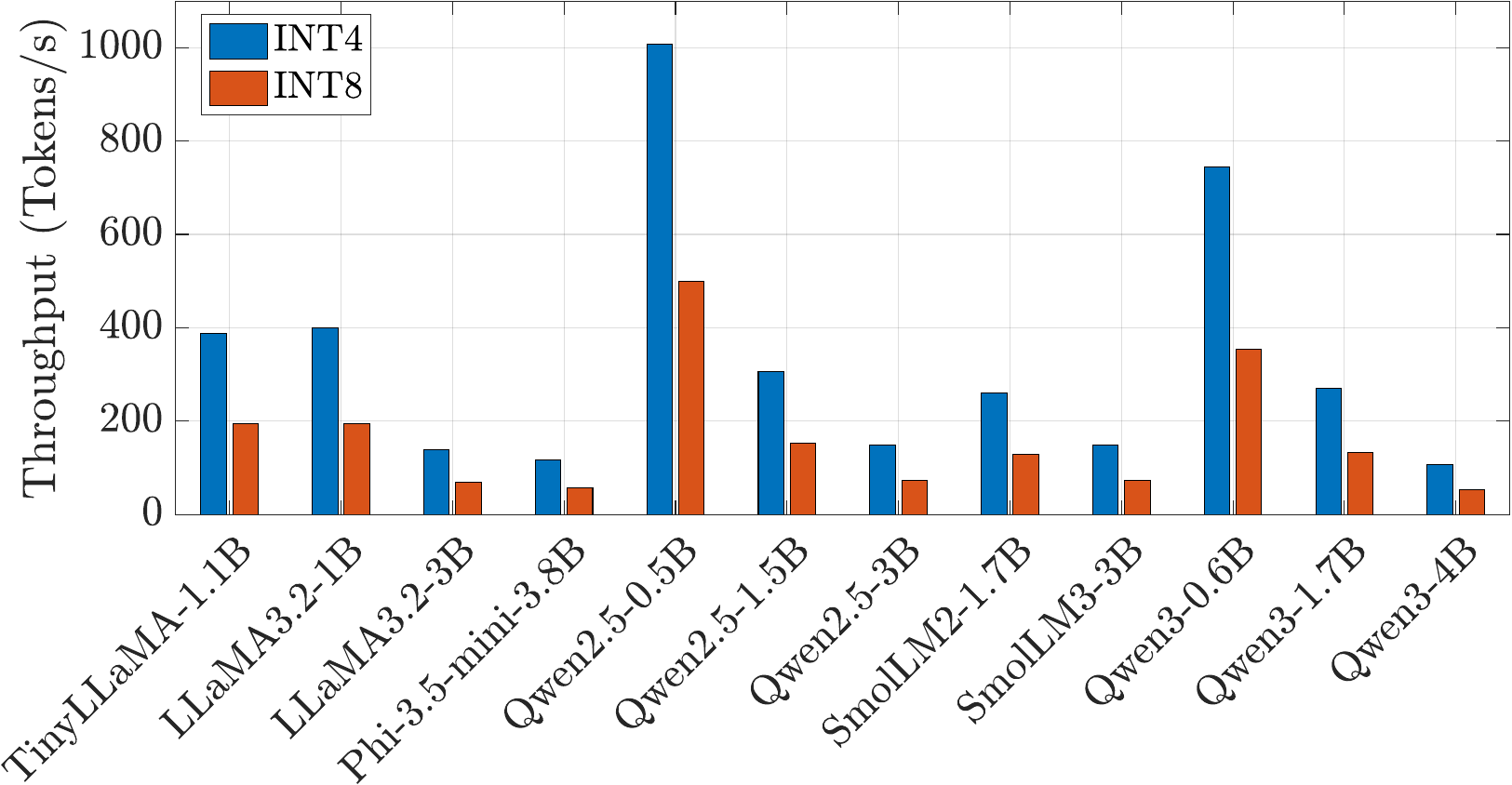}}\hfill
    \subfloat[]{\includegraphics[width=0.95\columnwidth]{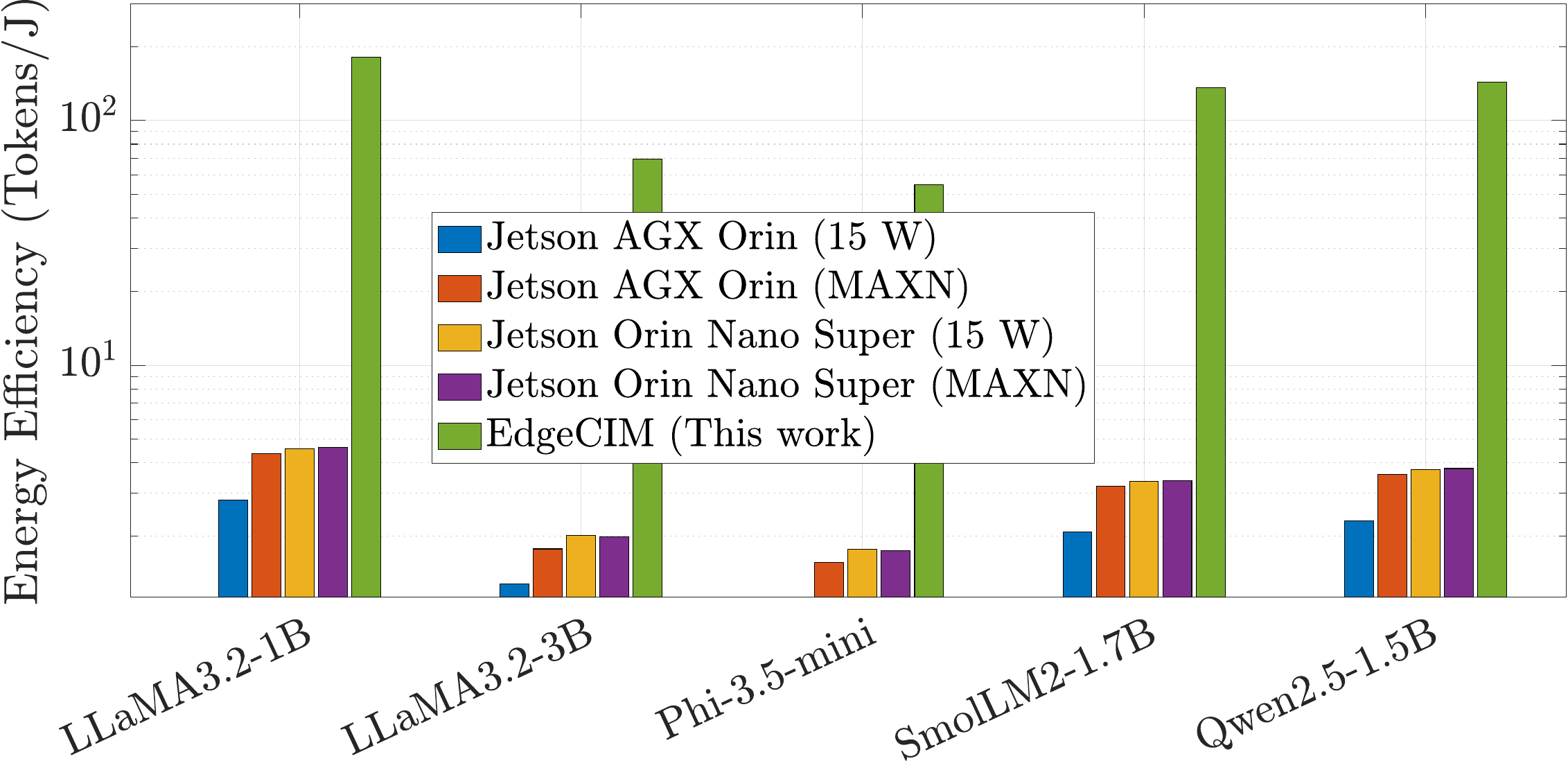}}\hfill
    \subfloat[]{\includegraphics[width=0.95\columnwidth]{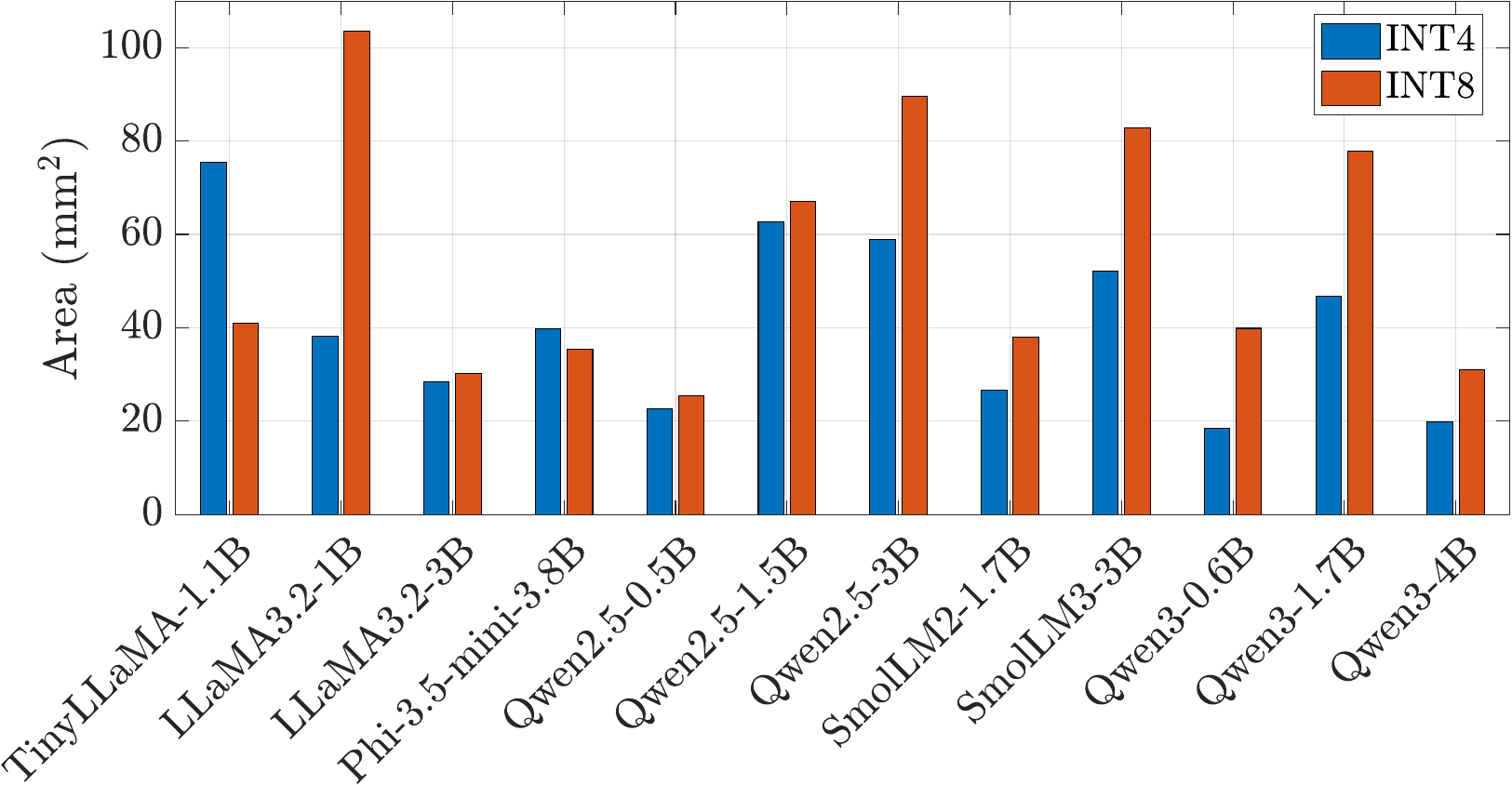}}
    \caption{(a) Throughput, (b) energy efficiency, and (c) area across SLMs for INT4 and INT8 precision at $\alpha = 1$.}
    \label{fig:results}
\end{figure}

\subsection{Performance Evaluation}
\label{sec:evaluation}
We then evaluate the end-to-end decoding performance of the resulting DSE-selected configurations. Fig. \ref{fig:results}(a) and (b) report throughput and energy efficiency, respectively, for all models at INT4 and INT8 precision. The results show that the proposed accelerator achieves high throughput and energy efficiency across all workloads. For instance, at INT4 precision, Qwen2.5-0.5B reaches over 1000 tokens/s with an efficiency exceeding 600 tokens/J, while TinyLLaMA-1.1B delivers nearly 400 tokens/s at more than 120 tokens/J. Even for larger models such as SmolLM3-3B, the accelerator sustains 148.7 tokens/s with around 72.5 tokens/J, highlighting its ability to maintain competitive efficiency as model size increases. Precision also plays a significant role, with the move from INT8 to INT4 nearly doubling throughput and energy efficiency across models. Although the optimization objective is latency only ($\alpha = 1$), the accelerator continues to deliver strong energy efficiency, confirming its suitability for edge deployment of compact language models where both metrics are critical. {Fig.~\ref{fig:results}(c) shows that the DSE-selected hardware configurations occupy 18.4 to 103.6~mm\textsuperscript{2}, remaining within the 105~mm\textsuperscript{2} reference area budget. Across the evaluated SLMs, processing elements account for an average of 62.93\% of the modeled accelerator area, followed by on-chip SRAM buffers at 25.79\%. Adder trees, accumulators, and dedicated functional units contribute 8.21\%, while control and interconnect-related logic account for the remaining 3.07\%. The reported accelerator area excludes the DRAM controller associated with the external LPDDR5X memory subsystem.}

{It is worth noting that, in some cases, the INT4 configuration exhibits a larger area than the corresponding INT8 configuration because the DSE independently optimizes the hardware configuration for each precision under the latency-only objective ($\alpha=1$). For TinyLLaMA-1.1B, the DSE selects 16 clusters, 32 tiles per cluster, and 4 PEs per tile for INT4, compared with 8 clusters, 8 tiles per cluster, and 16 PEs per tile for INT8. For Phi-3.5-mini-3.8B, INT4 uses 12 clusters, 24 tiles per cluster, and 4 PEs per tile, whereas INT8 uses 6 clusters, 16 tiles per cluster, and 9 PEs per tile. Thus, the larger INT4 area in these cases results from the DSE selecting a more parallel hardware configuration.}

{To further assess the effectiveness and reproducibility of
the GA-based DSE, we compare GA against random search
across the 12 evaluated SLMs at INT4 and INT8 precision. Both methods
explore the same hardware design space using 1,000 candidate
evaluations per run and five independent random seeds per
configuration. Using the latency-only objective, GA achieves lower mean best-found latency
than random search across all 24 workload--precision configurations, with reductions ranging from 8\% to 21\%.}

{While Fig.~\ref{fig:results} reports the performance obtained through application-specific optimization, where the DSE independently selects the hardware configuration for each model and precision, some deployments may instead require a single accelerator capable of supporting a diverse workload. To evaluate this scenario, we additionally perform a joint DSE over the complete workload suite consisting of all evaluated SLMs under both INT4 and INT8 precisions. For each candidate hardware configuration, the fitness is computed as the geometric mean of the latency-energy objective across all workloads. For the latency-optimized setting ($\alpha=1$), this corresponds to minimizing the geometric mean latency over the entire workload suite. {The resulting shared hardware configuration consists of $4\times2$ clusters, $2\times4$ tiles per cluster, $T_A=4$ active tiles per cluster, 16 PEs per tile, and 4096-bit interconnect buses, occupying a fixed area of 44.66~mm\textsuperscript{2}.} Under this common architecture, EdgeCIM achieves a geometric mean latency of 0.73~s across all 24 workloads (12 models $\times$ 2 precisions), while sustaining throughput ranging from 49 to 900~tokens/s and energy efficiency ranging from 20.6 to 491~tokens/J. Compared with the application-specific optimization results in Fig.~\ref{fig:results}, where the throughput ranges from 52.2 to 1007.9~tokens/s and the energy efficiency ranges from 24.7 to 603.8~tokens/J, the shared configuration preserves most of the achievable performance while requiring only a single hardware implementation. These results demonstrate that, although application-specific optimization provides the highest achievable performance for individual models, EdgeCIM also maintains competitive performance when a single hardware configuration is shared across a diverse set of SLM workloads.}

\subsection{Comparison with commercial edge GPUs and NPUs}
\label{sec:gpu_comparison}
{We then compare EdgeCIM against commercial edge GPUs using the NVIDIA Jetson AGX Orin and Jetson Orin Nano Super, which were directly benchmarked under matched inference settings. Both platforms were evaluated using INT4 quantization, a batch size of one, 128 prefill tokens, and 128 generated tokens, consistent with the EdgeCIM evaluation methodology. The comparison includes LLaMA3.2-1B, LLaMA3.2-3B, Phi-3.5-mini-3.8B, SmolLM2-1.7B, and Qwen2.5-1.5B. {Inference was performed using MLC LLM, following the benchmarking methodology of NVIDIA Jetson AI Lab~\cite{jetsonailab2025}. For each evaluated model, we used the corresponding \texttt{dusty-nv} \texttt{q4f16\_ft} checkpoint. Both Jetson platforms run NVIDIA Jetson Linux (L4T) R36.5.0, with CUDA 12.6, MLC LLM 0.1.3, and TVM 0.18.1. The MLC runtime uses the CUDA backend targeting \texttt{sm\_87}, with CUDA Graphs and CUTLASS enabled.}

{For each model, one warm-up inference is performed to mitigate initialization effects and is excluded from the reported results. Four prompts are then executed sequentially, and the decoding throughput (tokens/s) is averaged across the four runs.} {Power is monitored using NVIDIA \texttt{tegrastats}, with the sampling interval configured to 100~ms. A 5~s idle baseline is recorded before workload execution. For the AGX Orin, following NVIDIA's power-monitoring guidance~\cite{nvidia_agx_orin_power}, module-level power is estimated by summing the \texttt{VDD\_GPU\_SOC}, \texttt{VDD\_CPU\_CV}, and \texttt{VIN\_SYS\_5V0} monitored rails. For the Orin Nano Super, module-level power is obtained from the \texttt{VDD\_IN} rail.} {To isolate power consumption during decoding, the decoding interval for each run is identified using the timestamps of the first and last streamed tokens. For each monitored rail, power is averaged over this interval, and the corresponding idle average is subtracted to obtain the dynamic decoding power. Energy efficiency is calculated as decoding throughput divided by the idle-subtracted dynamic decoding power.}

% Inference was performed using the NVIDIA Jetson AI Lab MLC LLM benchmark \cite{jetsonailab2025} configured with the above inference settings. Throughput was obtained from the benchmark-reported average decoding throughput, while energy efficiency was computed from the corresponding measured decoding power after subtracting the idle power.}

\begin{figure}[!b]
    \centering
    \subfloat[]{\includegraphics[width=0.95\columnwidth]{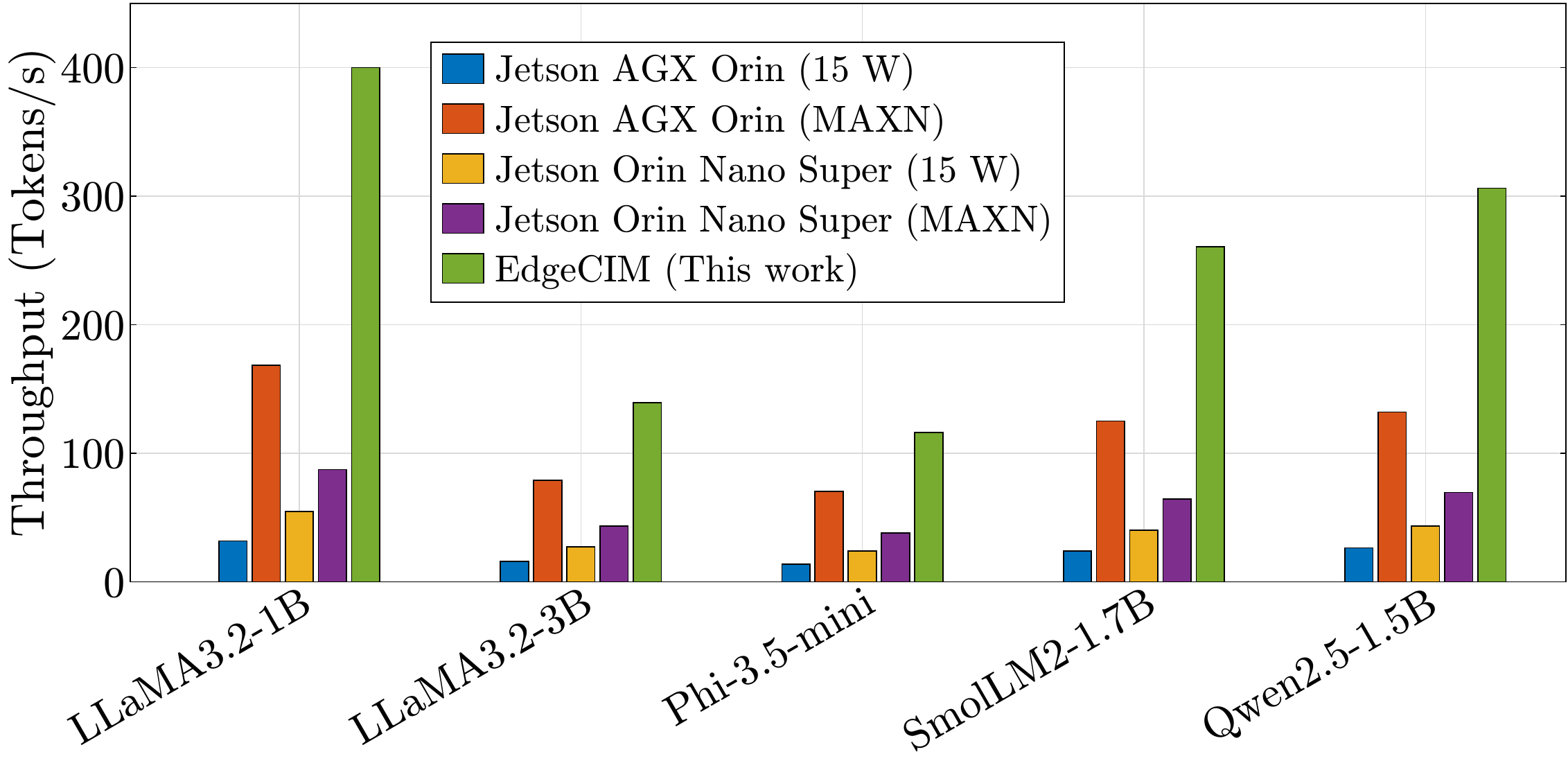}}\hfill
    \subfloat[]{\includegraphics[width=0.95\columnwidth]{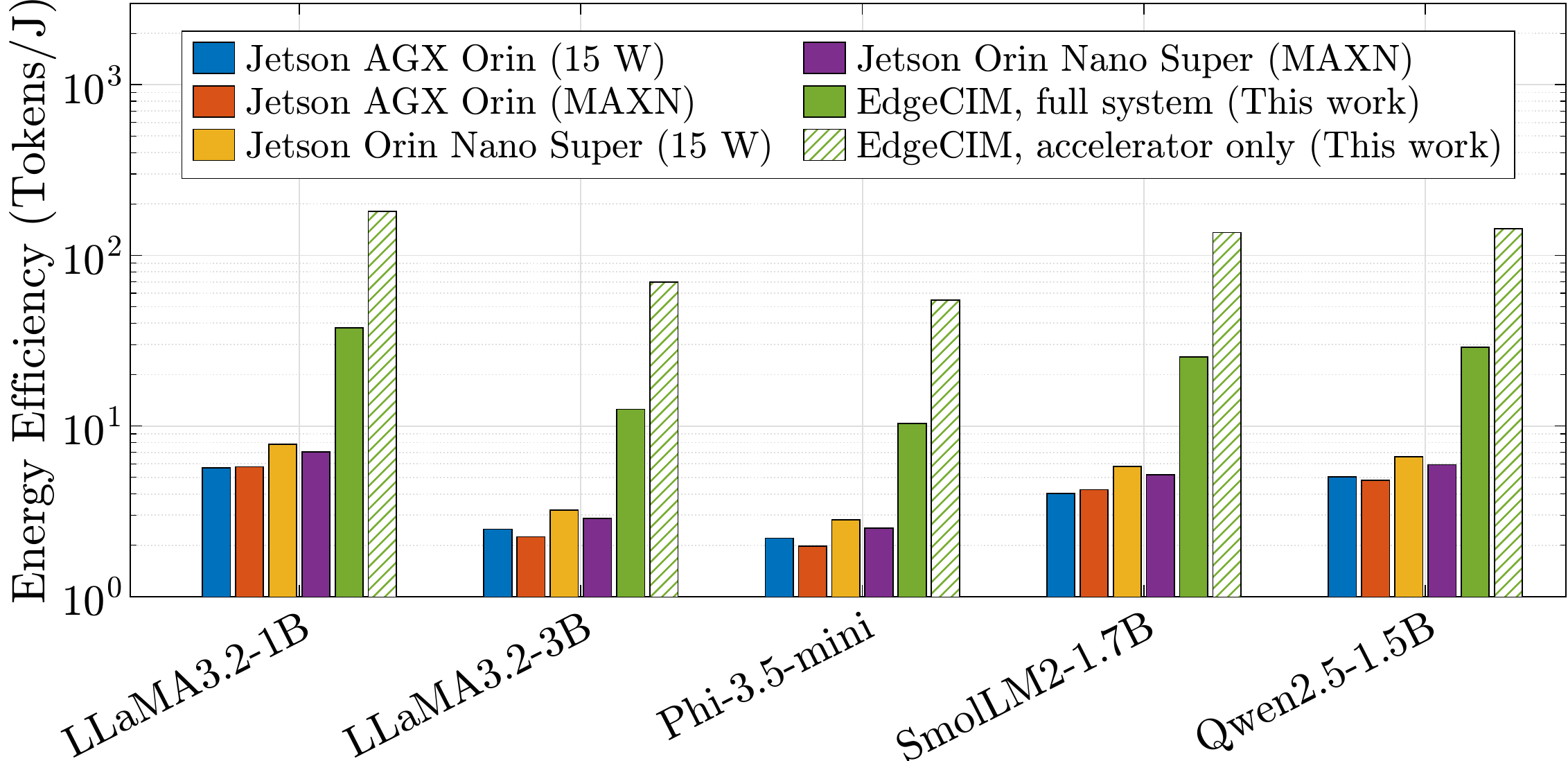}}
        \caption{{Comparison of EdgeCIM and NVIDIA Jetson edge GPUs at INT4 precision: (a) throughput and (b) energy efficiency.}}
    \label{fig:gpu_comp}
\end{figure}

Fig.~\ref{fig:gpu_comp}(a) and (b) compare the throughput and energy efficiency, respectively, of EdgeCIM with those of the NVIDIA Jetson AGX Orin and Jetson Orin Nano Super platforms. For the Jetson platforms, results are reported under both the 15~W low-power and MAXN operating modes. Across all evaluated models, EdgeCIM consistently outperforms both commercial GPU platforms. Under the 15~W operating mode, EdgeCIM achieves 8.4$\times$ to 12.5$\times$ higher throughput than the Jetson AGX Orin and 4.8$\times$ to 7.3$\times$ higher throughput than the Jetson Orin Nano Super. Even when compared against the highest-performance MAXN configuration, EdgeCIM still provides throughput improvements of 1.7$\times$ to 2.4$\times$ over the Jetson AGX Orin and 3.1$\times$ to 4.6$\times$ over the Jetson Orin Nano Super. {In terms of energy efficiency, we report two estimates for EdgeCIM: accelerator-only and accelerator with off-chip DRAM. The accelerator-only estimate considers the modeled on-chip energy consumption, while the accelerator-with-DRAM estimate additionally accounts for external LPDDR5X DRAM energy consumption~\cite{micron_lpddr5x_2022}. Across the evaluated SLMs, EdgeCIM achieves an estimated energy efficiency of 54.79 to 181.05~tokens/J for the accelerator-only configuration and 10.34 to 37.67~tokens/J when including external DRAM energy. In comparison, the Jetson platforms achieve 1.98 to 7.83~tokens/J across both operating modes, based on measured module-level dynamic power. Although the energy-efficiency results reflect different power-accounting boundaries, with EdgeCIM estimates covering accelerator and external DRAM energy consumption and Jetson measurements reflecting module-level dynamic power, including CPU and SoC components, they highlight the potential of EdgeCIM for energy-efficient SLM decoding at the edge.}

% Energy-efficiency improvements are even more pronounced. Compared with the 15~W operating mode, EdgeCIM achieves $24.9\times$--$33.7\times$ higher energy efficiency than the Jetson AGX Orin and $19.4\times$--$23.5\times$ higher energy efficiency than the Jetson Orin Nano Super. Under the MAXN operating mode, the corresponding improvements are $27.7\times$--$32.1\times$ and $21.7\times$--$26.2\times$, respectively, as the higher throughput of MAXN is obtained at a proportionally higher power consumption.}

% {To provide additional context for these results, we also compare the corresponding power consumption. Across the evaluated workloads, EdgeCIM operates within a power range of $1.92$--$2.21$~W, while the evaluated Jetson platforms span $5.27$--$8.49$~W under the constrained 15~W operating mode and $11.72$--$35.59$~W under MAXN. Although the Jetson measurements represent complete system power, including the host CPU, whereas the EdgeCIM values correspond only to the modeled accelerator, these results indicate that EdgeCIM achieves substantially higher throughput and energy efficiency while operating within practical edge power budgets, rather than relying on increased power provisioning.}

{Table~\ref{tab:throughput} further extends the comparison to additional commercial edge processors by comparing the throughput of EdgeCIM for LLaMA3.2-3B (INT4) against the Snapdragon 8 Elite Mobile, Snapdragon X Elite, and Snapdragon X2 Elite. The reported results for these platforms are taken from~\cite{qualcomm} and use INT4 quantization with a 128-token prefill, matching the principal inference settings adopted in our evaluation. Accordingly, they are presented as reference points under closely matched reported settings rather than as fully controlled platform-to-platform comparisons. To provide the strongest commercial GPU baseline, Table~\ref{tab:throughput} reports only the measured MAXN results for the Jetson platforms. Under this comparison, EdgeCIM achieves a throughput of 139.3~tokens/s for LLaMA3.2-3B (INT4), corresponding to improvements of 1.76$\times$, 3.21$\times$, 4.64$\times$, 7.07$\times$, and 3.22$\times$ over the Jetson AGX Orin, Jetson Orin Nano Super, Snapdragon 8 Elite Mobile, Snapdragon X Elite, and Snapdragon X2 Elite, respectively.}

\begin{table*}[!t]
\centering
\footnotesize
\setlength{\tabcolsep}{4pt}    
\renewcommand{\arraystretch}{1.05} 
\caption{Comparison of EdgeCIM with SOTA CIM accelerators.}
\begin{tabularx}{\textwidth}{l l l c l Y c}
\toprule
\textbf{Work} & \textbf{CIM Technology} & \textbf{Stage(s) Optimized For} 
& \textbf{End-to-End Optimization} & \textbf{Model Supported} 
& \textbf{Precision} & \textbf{TOPS/W/mm\textsuperscript{2}} \\
\midrule
\textbf{EdgeCIM (this work)} 
& CMOS  
& Projection+Attention+FFN 
& \cmark 
& Decoder-only 
& 4/8 bits 
& 7.03\textsuperscript{$*$} \\
X-Former~\cite{sridharan2023xformer} 
& CMOS + ReRAM 
& Projection+Attention 
& \xmark 
& Encoder-only 
& 8 bits  
& NA\textsuperscript{$\dagger$} \\
TranCIM~\cite{tu2023trancim} 
& CMOS 
& Attention+FFN 
& \xmark  
& Encoder-only 
& 8+16 bits  
& 3.06 \\
ReTransformer~\cite{yang2020retransformer} 
& ReRAM 
& Attention 
& \xmark 
& Encoder-decoder 
& 8 bits 
& NA\textsuperscript{$\ddagger$} \\
iMTransformer~\cite{laguna2022hardware} 
& CMOS + FeFET 
& Attention 
& \xmark 
& Encoder-decoder 
& 8 bits 
& 1.64 \\
\bottomrule
\end{tabularx}
\label{tab:cim-compare}
\footnotesize
\noindent\textsuperscript{$*$}For $h^*$: $C_v{=}2$, $C_h{=}3$, $T_{total}{=}16$, $P^2{=}16$. \\ \textsuperscript{$\dagger$}X-Former reports only 467.68 GOPS/W;\;
\textsuperscript{$\ddagger$}ReTransformer reports only 13440 GOPS/W.
\end{table*}

\begin{table}[!t]
\centering
\small
\caption{Throughput of LLaMA3.2-3B (INT4) on edge platforms. {The Jetson entries are our measured MAXN results; the Qualcomm Snapdragon entries are taken from \cite{qualcomm}.}}
\label{tab:throughput}
\begin{tabular}{lr}
\toprule
\textbf{Platform} & \textbf{Throughput (tokens/s)} \\
\midrule
Jetson Orin Nano Super (MAXN) & 43.4 \\
Jetson AGX Orin (MAXN)        & 79.1 \\
Snapdragon 8 Elite Mobile     & 30 \\
Snapdragon X Elite            & 19.7 \\
Snapdragon X2 Elite           & 43.2 \\
\textbf{EdgeCIM (This work)}  & \textbf{139.3} \\
\bottomrule
\end{tabular}
\end{table}

% \subsection{Comparison with CIM-based accelerators}
% We next compare EdgeCIM against prior CIM-based accelerators. Existing CIM accelerators target GEMM-dominated kernels, such as those in encoder-style models, rather than the GEMV-heavy decoding pipeline. \textit{X-Former} \cite{sridharan2023xformer} uses ReRAM crossbars with CMOS logic to accelerate projection and attention kernels, but is evaluated only on encoder-style models such as BERT. \textit{TranCIM} \cite{tu2023trancim} employs a digital bitline-transpose CIM macro with reconfigurable parallel/pipeline modes (for attention/FFN respectively) and a sparse attention scheduler, yet its efficiency gains are limited to encoder prefill (matrix-matrix multiplications). \textit{ReTransformer} \cite{yang2020retransformer} and \textit{iMTransformer} \cite{laguna2022hardware} similarly focus on scaled dot-product attention in encoder-decoder settings using ReRAM and CMOS+FeFET arrays, without distinguishing prefill from decoding or optimizing the decode stage. These works emphasize isolated attention kernel, without tackling the \emph{end-to-end optimization} needed for decoder-only LMs. 

\subsection{Comparison with CIM-based accelerators}

Finally, we compare EdgeCIM with prior CIM-based Transformer accelerators. Although prior designs can execute masked attention or batch-size-one workloads, their published architectures and mappings are not specifically optimized for end-to-end autoregressive decoding, where only one new token becomes available per generation step, projection and FFN layers are dominated by GEMVs, model weights must be repeatedly accessed, and the KV-cache grows with sequence length.

X-Former~\cite{sridharan2023xformer} uses separate projection and attention engines and overlaps their execution across blocks of concurrently available input tokens. This sequence-block overlap cannot be directly exploited across future generated tokens, while its assumption of spatially resident static weights would need to be replaced by a streaming policy for edge-scale SLMs. TranCIM~\cite{tu2023trancim} supports layer-dependent pipeline and parallel modes, but its published evaluation uses a BERT-oriented full-sequence mapping and sparse-attention scheduler. ReTransformer~\cite{yang2020retransformer} and iMTransformer~\cite{laguna2022hardware} primarily optimize attention computation and its dynamic operands rather than jointly mapping the complete decoder token path. Efficiently adapting these designs would therefore require substantial changes to support per-token projection and FFN GEMVs, growing KV-cache traffic, GQA-based key/value reuse, streamed model weights, and stage-dependent resource allocation under constrained edge-memory bandwidth.

EdgeCIM addresses these requirements using a unified, stage-reconfigurable SRAM-CIM architecture. In particular, the number of active tiles, $T_A$, is optimized independently from the total physical tile capacity, allowing the execution width, streamed weight-partition size, and communication-computation overlap to be jointly configured for projection, attention, output projection, and FFN stages. Thus, EdgeCIM's distinction is not simply its support for decoder-only models, but its bandwidth-aware, end-to-end hardware--software co-optimization for batch-size-one autoregressive inference.

For the quantitative comparison, autoregressive throughput in tokens/s is not reported by X-Former, TranCIM, ReTransformer, or iMTransformer and cannot be reconstructed reliably from their published latency or arithmetic-throughput results. Tokens/s depends on the evaluated model, batch size, prompt and generation lengths, KV-cache organization, memory behavior, and execution of the complete decoder stack. In particular, full-sequence encoder latency cannot be converted directly into decoding throughput because encoder tokens are concurrently available, whereas autoregressive generation produces one causally dependent token per decoding step. Likewise, converting TOPS into tokens/s would require the achieved utilization, operation count, memory traffic, communication overhead, and non-CIM latency for every decoder stage, which are not reported consistently by these works. We therefore provide direct tokens/s comparisons against decoder-capable edge platforms under matched batch-size-one inference conditions, while Table~\ref{tab:cim-compare} compares prior CIM designs in terms of technology, supported stages, model type, precision, and normalized efficiency, including TOPS/W/mm\textsuperscript{2}.

The reported TOPS/W/mm\textsuperscript{2} values provide a complementary comparison of compute efficiency and density, but they should not be interpreted as end-to-end generation-throughput measurements. The values follow the respective evaluation methodologies of the individual works and may differ in process technology, precision, optimized workload coverage, operation-counting convention, and the treatment of peripheral and memory-system overheads. Nevertheless, under their reported configurations, TranCIM and iMTransformer achieve 3.06 and 1.64 TOPS/W/mm\textsuperscript{2}, respectively, whereas EdgeCIM achieves 7.03 TOPS/W/mm\textsuperscript{2} while supporting the complete decoder-only token path. This result demonstrates that EdgeCIM combines higher reported normalized compute efficiency with end-to-end support for projection, attention, output projection, and FFN execution.

\section{Conclusion}
\label{sec:conclusion} 
{In this work, we presented EdgeCIM, a hardware--software co-design framework for accelerating latency-critical, batch-size-one decoder-only inference of SLMs on edge devices.} The framework integrates a genetic algorithm-based optimization process with an analytical simulator to explore the CIM hardware design space and identify optimal configurations. At the architectural level, EdgeCIM employs a tiled hierarchy of digital SRAM-based CIM macros and introduces an active-tile pipelined mapping strategy to optimize performance. At INT4 precision, the accelerator sustains an average of 336.4 tokens/s and 173 tokens/J across multiple SLM benchmarks, confirming its ability to balance high performance with energy efficiency under edge constraints. {One promising direction is to extend EdgeCIM into a heterogeneous
architecture that combines the proposed CIM fabric for GEMV-dominated
autoregressive decoding with a complementary GPU/NPU/TPU-style compute
tier optimized for GEMM-intensive workloads, including the prefill
phase and batched inference.}

% One promising direction is to extend EdgeCIM into a heterogeneous architecture that combines the proposed CIM fabric for GEMV-dominated operations with a complementary GPU/NPU/TPU-style compute tier optimized for the increasingly GEMM-intensive workloads arising under batched inference.}

\section*{Acknowledgment}
{The authors used OpenAI's ChatGPT during the preparation of this manuscript to assist with improving the readability, grammar, and clarity of portions of the text. All AI-generated content was carefully reviewed, edited, and verified by the authors, who take full responsibility for the final manuscript. The AI tool was not used to generate the scientific contributions, experimental results, analyses, or conclusions presented in this work.}

\begin{IEEEbiography}
[{\includegraphics[width=1in,height=1.25in]{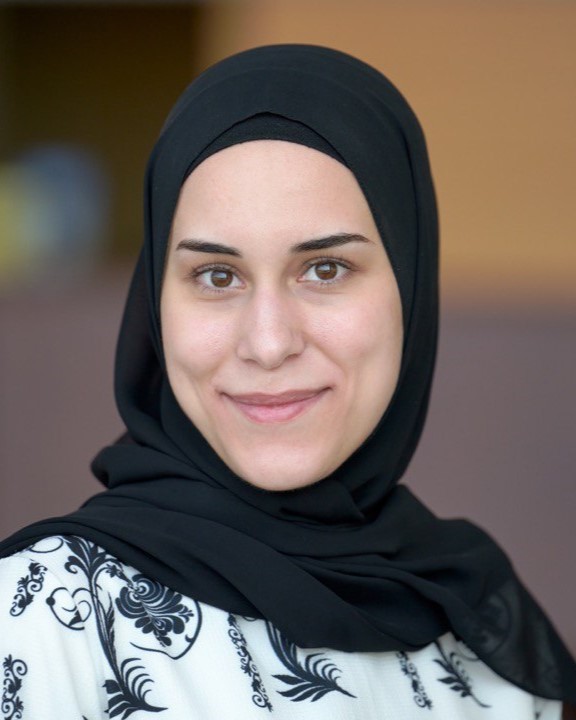}}]
{Jinane Bazzi} received the B.E. degree (with High Distinction) in computer and communications engineering from the American University of Beirut (AUB), Lebanon, in 2021, and the M.Sc. degree in electrical and computer engineering from the King Abdullah University of Science and Technology (KAUST), Saudi Arabia, in 2023, where she is currently pursuing the Ph.D. degree. Her research interests include hardware--software co-design for compute-in-memory architectures and AI hardware acceleration. She received the IEEE International Conference on Microelectronics (ICM) Best Paper Award in 2020.
\end{IEEEbiography}

\begin{IEEEbiography}
[{\includegraphics[width=1in,height=1.25in]{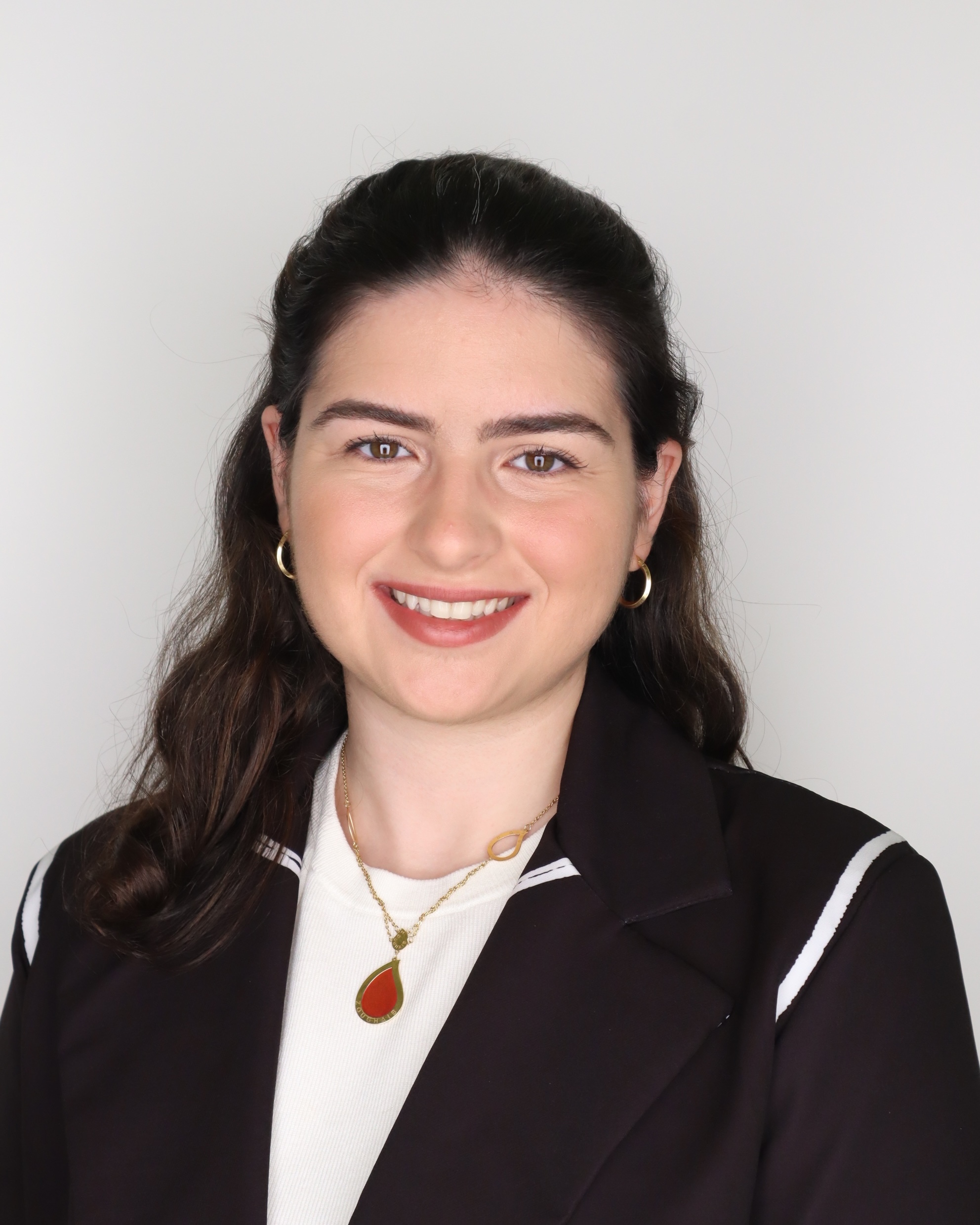}}]
{Mariam Rakka} received the B.E. degree (with High Distinction) in computer and communications engineering with two minors in mathematics and business administration from the American University of Beirut in 2020, and the M.S. degree in electrical and computer engineering in 2022 from the University of California, Irvine, where she completed her PhD. She has published work that focuses on in-memory computing technologies, hardware accelerators, rare fail event estimation, and hardware-friendly neural networks. She joined Arm Inc. as a research intern during the summers of 2022, 2023, and 2024.
\end{IEEEbiography}

\begin{IEEEbiography}
[{\includegraphics[width=1in,height=1.1in]{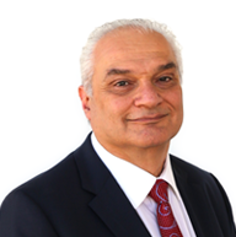}}]
{Fadi Kurdahi} received the BE degree in electrical engineering from the American
University of Beirut in 1981, and the PhD degree from the University of Southern California in 1987. Since then, he has been a faculty with the Electrical Engineering and Computer Science Department, University of California at Irvine, where he was the
associate dean for Graduate and Professional Studies with the Samueli School of Engineering during 2017-2022, and is currently the director of the Center for
Embedded \& Cyber-physical Systems, comprised of world-class researchers in the general area of embedded and cyber-physical
systems. He is a fellow of the AAAS. He was the recipient of the best paper awards of IEEE Transactions on Very Large Scale Integration in 2002 and of ISQED at 2006, and four other distinguished paper awards at DAC, EuroDAC, ASP-DAC, and ISQED.
\end{IEEEbiography}

\begin{IEEEbiography}
[{\includegraphics[width=1in,height=1.25in]{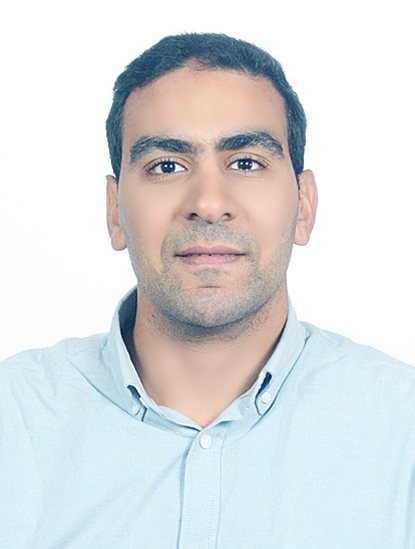}}]
{Mohammed E. Fouda} received a B.Sc. degree (Hons.) in Electronics and Communications Engineering and an M.Sc. degree in Engineering Mathematics from the Faculty of Engineering, Cairo University, Cairo, Egypt, in 2011 and 2014, respectively. Dr. Fouda received a Ph.D. degree from the University of California, Irvine, USA, in 2020. Currently, he is an applied research lead at Rain Neuromorphics Inc., USA. Dr. Fouda has published more than 220 peer-reviewed journal and conference papers. His H-index is 39, and he has been cited more than 5100 times. His research interests include AI hardware and algorithms, Agentic AI, and AI safety and Security. He serves as a peer reviewer for many prestigious journals and conferences. He also serves as an associate editor in many journals and a technical program committee member at many conferences. 
\end{IEEEbiography}

\begin{IEEEbiography}
[{\includegraphics[width=1in,height=1.25in]{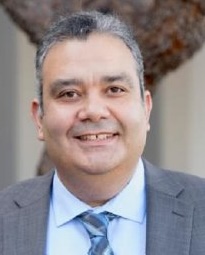}}]
{Ahmed Eltawil} received the bachelor's and master's degrees from Cairo University, in 1997 and 1999, respectively, and the Doctorate degree from the University of California, Los Angeles, in 2003. He is currently a Professor and an Associate Dean for Research with the Computer, Electrical, and Mathematical Sciences and Engineering (CEMSE) Division, King Abdullah University of Science and Technology (KAUST). Previously, he was a Professor in electrical engineering and computer science with the University of California, Irvine (UCI), from 2005 to 2021. At KAUST, he established the Communication and Computing Systems Laboratory (CCSL) to conduct research on efficient architectures for computing and communications systems, with a particular focus on mobile wireless systems. His research interests include low-power mobile systems, machine learning platforms, sensor networks, body area networks, and critical infrastructure networks. He served as a Distinguished Lecturer for IEEE ComSoc in 2023 and 2024. Additionally, he holds senior membership with the National Academy of Inventors, USA. He received several recognitions and awards, including the U.S. National Science Foundation CAREER Award, the 2021 ``Innovator of the Year'' Award by the Henry Samueli School of Engineering, University of California, Irvine, and two U.S. Congress Certificates of Merit, among other recognitions. He has served in numerous editorial roles over the years and as an expert reviewer for national and international funding agencies and review boards.
\end{IEEEbiography}

\bibliographystyle{IEEEtran}
\bibliography{sample-base}

@manual{hspice,
  title        = {{HSPICE} Reference Manual: Commands and Control Options},
  organization = {Synopsys, Inc.},
  address      = {Mountain View, CA},
}

@manual{micron_lpddr5x_2022,
  author       = {{Micron Technology, Inc.}},
  title        = {{Automotive LPDDR5X SDRAM:
                   MT62F1G32D2, MT62F2G32D4, MT62F4G32D8}},
  organization = {Micron Technology, Inc.},
  year         = {2022},
  month        = nov,
  note         = {Rev. B, Technical Datasheet}
}

@misc{qualcomm,
  title        = {{Llama-v3.2-3B-Instruct}},
  author       = {{Qualcomm AI Hub}},
  howpublished = {\url{https://aihub.qualcomm.com/models/llama_v3_2_3b_instruct?searchTerm=llama-v3}},
  note = {Accessed: August 1, 2026},
}

@misc{jetsonailab2025,
  title        = {NVIDIA Jetson AI Lab},
author = {NVIDIA},
  howpublished = {\url{https://www.jetson-ai-lab.com/benchmarks.html}}
}

@inproceedings{chih202116,
  title={16.4 An 89TOPS/W and 16.3 TOPS/mm 2 all-digital SRAM-based full-precision compute-in memory macro in 22nm for machine-learning edge applications},
  author={Chih, Yu-Der and Lee, Po-Hao and Fujiwara, Hidehiro and Shih, Yi-Chun and Lee, Chia-Fu and Naous, Rawan and Chen, Yu-Lin and Lo, Chieh-Pu and Lu, Cheng-Han and Mori, Haruki and others},
  booktitle={2021 IEEE International Solid-State Circuits Conference (ISSCC)},
  volume={64},
  pages={252--254},
  year={2021},
  organization={IEEE}
}

@article{lin2024awq,
  title={Awq: Activation-aware weight quantization for on-device llm compression and acceleration},
  author={Lin, Ji and Tang, Jiaming and Tang, Haotian and Yang, Shang and Chen, Wei-Ming and Wang, Wei-Chen and Xiao, Guangxuan and Dang, Xingyu and Gan, Chuang and Han, Song},
  journal={Proceedings of machine learning and systems},
  volume={6},
  pages={87--100},
  year={2024}
}

@misc{nvidia_jetson_thor_datasheet,
  author       = {{NVIDIA Corporation}},
  title        = {{NVIDIA Jetson Thor Series Modules Data Sheet}},
  year         = {2025},
  howpublished = {\url{https://developer.nvidia.com/downloads/assets/embedded/secure/jetson/thor/docs/jetson-thor-series-modules-datasheet_ds-11945-001.pdf}},
  note         = {Accessed: 2026-07-25}
}

@misc{nvidia_jetson_orin,
  author       = {{NVIDIA Corporation}},
  title        = {{NVIDIA Jetson AGX Orin for Next-Generation Robotics}},
  year         = {2026},
  howpublished = {\url{https://www.nvidia.com/en-us/autonomous-machines/embedded-systems/jetson-orin/}},
  note         = {Accessed: 2026-07-25}
}

@misc{qualcomm_snapdragon_x_elite,
  author       = {{Qualcomm Technologies, Inc.}},
  title        = {{Snapdragon X Elite Platform}},
  year         = {2026},
  howpublished = {\url{https://www.qualcomm.com/laptops/products/snapdragon-x-elite}},
  note         = {Accessed: 2026-07-25}
}

@article{zheng2026empirical,
  title={An empirical study of qwen3 quantization},
  author={Zheng, Xingyu and Li, Yuye and Chu, Haoran and Feng, Yue and Ma, Xudong and Wang, Zining and Luo, Jie and Guo, Jinyang and Qin, Haotong and Magno, Michele and others},
  journal={Visual Intelligence},
  volume={4},
  number={1},
  pages={11},
  year={2026},
  publisher={Springer}
}

@misc{meta2024llama32,
  author       = {{Meta AI}},
  title        = {Llama\,3.2: Revolutionizing Edge AI and Vision with Open, Customizable Models},
  howpublished = {\url{https://ai.meta.com/blog/llama-3-2-connect-2024-vision-edge-mobile-devices/}},
  month        = sep,
  year         = {2024},
}

@misc{smollm3blog2025,
  author       = {Elie Bakouch and Loubna Ben Allal and Anton Lozhkov and et al.},
  title        = {SmolLM3: smol, multilingual, long-context reasoner},
  howpublished = {\url{https://huggingface.co/blog/smollm3}},
  month        = jul,
  year         = {2025},
}

@article{allal2025smollm2,
  title={SmolLM2: When Smol Goes Big--Data-Centric Training of a Small Language Model},
  author={Allal, Loubna Ben and Lozhkov, Anton and Bakouch, Elie and Bl{\'a}zquez, Gabriel Mart{\'\i}n and Penedo, Guilherme and Tunstall, Lewis and Marafioti, Andr{\'e}s and Kydl{\'\i}{\v{c}}ek, Hynek and Lajar{\'\i}n, Agust{\'\i}n Piqueres and Srivastav, Vaibhav and others},
  journal={arXiv preprint arXiv:2502.02737},
  year={2025}
}

@inproceedings{bazzi2026synthesizable,
  title={A Synthesizable Mixed-Precision DCIM Macro with Parallel Write and Compute},
  author={Bazzi, Jinane and Fouda, Mohammed E and Eltawil, Ahmed M},
  booktitle={2026 IEEE International Symposium on Circuits and Systems (ISCAS)},
  pages={1919--1923},
  year={2026},
  organization={IEEE}
}

@article{yang2025qwen3,
  author       = {An Yang and Anfeng Li and Baosong Yang and Beichen Zhang and Binyuan Hui and et al.},
  title        = {Qwen3 Technical Report},
  year         = {2025},
  journal={arXiv preprint arXiv:2505.09388}

}

@article{qwen2025qwen25technicalreport,
  author       = {An Yang and Baosong Yang and Beichen Zhang and Binyuan Hui and Bo Zheng and et al.},
  title        = {Qwen2.5 Technical Report},
  year         = {2025},
  journal={arXiv preprint arXiv:2412.15115}
}

@article{abdin2024phi3technicalreporthighly,
  author       = {M. Abdin and J. Aneja and H. Awadalla and A. Awadallah and A. Awan and et al.},
  title        = {Phi-3 Technical Report: A Highly Capable Language Model Locally on Your Phone},
  year         = {2024},
  journal={arXiv preprint arXiv:2404.14219}
}

@article{mesh,
  title={Domain-specific hardware accelerators},
  author={Dally, William J and Turakhia, Yatish and Han, Song},
  journal={Communications of the ACM},
  volume={63},
  number={7},
  pages={48--57},
  year={2020},
  publisher={ACM New York, NY, USA}
}

@inproceedings{yang2020retransformer,
  title={ReTransformer: ReRAM-based processing-in-memory architecture for transformer acceleration},
  author={Yang, Xiaoxuan and Yan, Bonan and Li, Hai and Chen, Yiran},
  booktitle={Proceedings of the 39th International Conference on Computer-Aided Design},
  pages={1--9},
  year={2020}
}

@inproceedings{cacti,
  title={Optimizing NUCA organizations and wiring alternatives for large caches with CACTI 6.0},
  author={Muralimanohar, Naveen and Balasubramonian, Rajeev and Jouppi, Norm},
  booktitle={40th Annual IEEE/ACM International Symposium on Microarchitecture (MICRO 2007)},
  pages={3--14},
  year={2007},
  organization={IEEE}
}

@article{zhang2024tinyllama,
  title={Tinyllama: An open-source small language model},
  author={Zhang, Peiyuan and Zeng, Guangtao and Wang, Tianduo and Lu, Wei},
  journal={arXiv preprint arXiv:2401.02385},
  year={2024}
}

@inproceedings{bazzi2024reconfigurable,
  title={Reconfigurable Precision SRAM-based Analog In-memory-compute Macro Design},
  author={Bazzi, Jinane and Jamil, Rachid and ElHajj, Dana and Kanj, Rouwaida and Fouda, Mohammed E and Eltawil, Ahmed},

booktitle={2024 IEEE International Symposium on Circuits and Systems (ISCAS)},
  pages={1--5},
  year={2024},
  organization={IEEE}
}

@inproceedings{bazzi2025reconfigurable,
  title={Reconfigurable Precision INT4-8/FP8 Digital Compute-in-Memory Macro for AI Acceleration},
  author={Bazzi, Jinane and Fouda, Mohammed E and Eltawil, Ahmed},
  booktitle={2025 IEEE International Symposium on Circuits and Systems (ISCAS)},
  pages={1--5},
  year={2025},
  organization={IEEE}
}

@article{kim2023full,
  title={Full stack optimization of transformer inference: a survey},
  author={Kim, Sehoon and Hooper, Coleman and Wattanawong, Thanakul and Kang, Minwoo and Yan, Ruohan and Genc, Hasan and Dinh, Grace and Huang, Qijing and Keutzer, Kurt and Mahoney, Michael W and others},
  journal={arXiv preprint arXiv:2302.14017},
  year={2023}
}

@inproceedings{hung20228,
  title={An 8-Mb DC-current-free binary-to-8b precision ReRAM nonvolatile computing-in-memory macro using time-space-readout with 1286.4-21.6 TOPS/W for edge-AI devices},
  author={Hung, Je-Min and Huang, Yen-Hsiang and Huang, Sheng-Po and Chang, Fu-Chun and Wen, Tai-Hao and Su, Chin-I and Khwa, Win-San and Lo, Chung-Chuan and Liu, Ren-Shuo and Hsieh, Chih-Cheng and others},
  booktitle={2022 IEEE International Solid-State Circuits Conference (ISSCC)},
  volume={65},
  year={2022},
  organization={IEEE}
}

@article{dao2022flashattention,
  title={Flashattention: Fast and memory-efficient exact attention with io-awareness},
  author={Dao, Tri and Fu, Dan and Ermon, Stefano and Rudra, Atri and R{\'e}, Christopher},
  journal={Advances in neural information processing systems},
  volume={35},
  pages={16344--16359},
  year={2022}
}

@inproceedings{xue202116,
  title={16.1 A 22nm 4Mb 8b-precision ReRAM computing-in-memory macro with 11.91 to 195.7 TOPS/W for tiny AI edge devices},
  author={Xue, Cheng-Xin and Hung, Je-Min and Kao, Hui-Yao and Huang, Yen-Hsiang and Huang, Sheng-Po and Chang, Fu-Chun and Chen, Peng and Liu, Ta-Wei and Jhang, Chuan-Jia and Su, Chin-I and others},
  booktitle={2021 IEEE International Solid-State Circuits Conference (ISSCC)},
  volume={64},
  pages={245--247},
  year={2021},
  organization={IEEE}
}

@inproceedings{chiu202222nm,
  title={A 22nm 4Mb STT-MRAM data-encrypted near-memory computation macro with a 192GB/s read-and-decryption bandwidth and 25.1-55.1 TOPS/W 8b MAC for AI operations},
  author={Chiu, Yen-Cheng and Yang, Chia-Sheng and Teng, Shih-Hsin and Huang, Hsiao-Yu and Chang, Fu-Chun and Wu, Yuan and Chien, Yu-An and Hsieh, Fang-Ling and Li, Chung-Yuan and Lin, Guan-Yi and others},
  booktitle={2022 IEEE International Solid-State Circuits Conference (ISSCC)},
  volume={65},
  pages={178--180},
  year={2022},
  organization={IEEE}
}

@article{liu2023mdcim,
  title={MDCIM: MRAM-Based Digital Computing-in-Memory Macro for Floating-Point Computation with High Energy Efficiency and Low Area Overhead},
  author={Liu, Liang and Tan, Lehao and Gan, Jie and Pan, Biao and Zhou, Jiahui and Li, Zhengliang},
  journal={Applied Sciences},
  volume={13},
  number={21},
  pages={11914},
  year={2023},
  publisher={MDPI}
}

@inproceedings{sarangi2021deepscaletool,
  title={Deepscaletool: A tool for the accurate estimation of technology scaling in the deep-submicron era},
  author={Sarangi, Satyabrata and Baas, Bevan},
  booktitle={2021 IEEE International Symposium on Circuits and Systems (ISCAS)},
  pages={1--5},
  year={2021},
  organization={Ieee}
}

@article{jssc22,
  title={ReDCIM: Reconfigurable digital computing-in-memory processor with unified FP/INT pipeline for cloud AI acceleration},
  author={Tu, Fengbin and Wang, Yiqi and Wu, Zihan and Liang, Ling and Ding, Yufei and Kim, Bongjin and Liu, Leibo and Wei, Shaojun and Xie, Yuan and Yin, Shouyi},
  journal={IEEE Journal of Solid-State Circuits},
  volume={58},
  number={1},
  pages={243--255},
  year={2022},
  publisher={IEEE}
}

@inproceedings{ankit2019puma,
  title={PUMA: A programmable ultra-efficient memristor-based accelerator for machine learning inference},
  author={Ankit, Aayush and Hajj, Izzat El and Chalamalasetti, Sai Rahul and Ndu, Geoffrey and Foltin, Martin and Williams, R Stanley and Faraboschi, Paolo and Hwu, Wen-mei W and Strachan, John Paul and Roy, Kaushik and others},
  booktitle={Proceedings of the Twenty-Fourth International Conference on Architectural Support for Programming Languages and Operating Systems},
  pages={715--731},
  year={2019}
}

@article{shafiee2016isaac,
  title={ISAAC: A convolutional neural network accelerator with in-situ analog arithmetic in crossbars},
  author={Shafiee, Ali and Nag, Anirban and Muralimanohar, Naveen and Balasubramonian, Rajeev and Strachan, John Paul and Hu, Miao and Williams, R Stanley and Srikumar, Vivek},
  journal={ACM SIGARCH Computer Architecture News},
  volume={44},
  number={3},
  pages={14--26},
  year={2016},
  publisher={ACM New York, NY, USA}
}

@inproceedings{jouppi2017datacenter,
  title={In-datacenter performance analysis of a tensor processing unit},
  author={Jouppi, Norman P and Young, Cliff and Patil, Nishant and Patterson, David and Agrawal, Gaurav and Bajwa, Raminder and Bates, Sarah and Bhatia, Suresh and Boden, Nan and Borchers, Al and others},
  booktitle={Proceedings of the 44th annual international symposium on computer architecture},
  pages={1--12},
  year={2017}
}

@article{boolq,
  title={BoolQ: Exploring the surprising difficulty of natural yes/no questions},
  author={Clark, Christopher and Lee, Kenton and Toutanova, Kristina},
  journal={arXiv preprint arXiv:1905.10044},
  year={2019}
}

@article{hellaswag,
  title={Hellaswag: Can a machine really finish your sentence?},
  author={Zellers, Rowan and Holtzman, Ari and others},
  journal={arXiv preprint arXiv:1905.07830},
  year={2019}
}

@article{winogrande,
  title={Winogrande: An adversarial winograd schema challenge at scale},
  author={Sakaguchi, Keisuke and Bras, Ronan Le and others},
  journal={Communications of the ACM},
  year={2021},
}

@article{zhou2025revisiting,
  title={Revisiting pruning vs quantization for small language models},
  author={Zhou, Zihan and Kurz, Simon and Zhao, Zhixue},
  journal={Findings of the Association for Computational Linguistics: EMNLP},
  volume={2025},
  pages={12055--12070},
  year={2025}
}

@article{brown2020language,
  title={Language models are few-shot learners},
  author={Brown, Tom B and Mann, Benjamin and Ryder, Nick and Subbiah, Melanie and others},
  journal={Advances in Neural Information Processing Systems},
  volume={33},
  pages={1877--1901},
  year={2020}
}

@inproceedings{vaswani2017attention,
  title={Attention is all you need},
  author={Vaswani, Ashish and Shazeer, Noam and Parmar, Niki and others},
  booktitle={Advances in Neural Information Processing Systems},
  pages={5998--6008},
  year={2017}
}

@inproceedings{devlin2019bert,
  title={BERT: Pre-training of deep bidirectional transformers for language understanding},
  author={Devlin, Jacob and Chang, Ming-Wei and Lee, Kenton and Toutanova, Kristina},
  booktitle={NAACL-HLT},
  pages={4171--4186},
  year={2019}
}

@article{kim2025hastily,
  title={HASTILY: Hardware-Software Co-Design for Accelerating Transformer Inference Leveraging Compute-in-Memory},
  author={Kim, Dong Eun and Sharma, Tanvi and Roy, Kaushik},
  journal={IEEE Transactions on Circuits and Systems for Artificial Intelligence},
  year={2025},
  publisher={IEEE}
}

@inproceedings{tambe2021edgebert,
  title={EdgeBERT: Sentence-level energy optimizations for latency-aware multi-task NLP inference},
  author={Tambe, Tushar and Haj-Ali, Ameer and others},
  booktitle={MICRO},
  pages={830--844},
  year={2021}
}

@misc{hu2023llamacpp,
  title={llama.cpp: A fast inference of LLaMA models},
  author={Georgi Gerganov and contributors},
  year={2023},
  howpublished={\url{https://github.com/ggerganov/llama.cpp}}
}

@article{verma2019imc,
  title={In-memory computing: Advances and prospects},
  author={Verma, Naveen and Shafiee, Ali and et al.},
  journal={IEEE Solid-State Circuits Magazine},
  volume={11},
  number={3},
  pages={43--55},
  year={2019}
}

@article{sridharan2023xformer,
  title={X-Former: In-memory acceleration of transformers},
  author={Sridharan, Shrihari and Stevens, Jacob R and Roy, Kaushik and Raghunathan, Anand},
  journal={IEEE Transactions on VLSI Systems},
  volume={31},
  number={8},
  pages={1223--1233},
  year={2023}
}

@article{tu2023trancim,
  title={TranCIM: Full-digital bitline-transpose CIM-based sparse transformer accelerator with pipeline/parallel reconfigurable modes},
  author={Tu, Fengbin and Wu, Zihan and Wang, Yiqi and Liang, Ling and Liu, Liu and Ding, Yufei and Liu, Leibo and Wei, Shaojun and Xie, Yuan and Yin, Shouyi},
  journal={IEEE Journal of Solid-State Circuits},
  volume={58},
  number={6},
  pages={1798--1809},
  year={2023}
}

@article{kim2019gpt_inference,
  title={Efficient inference for autoregressive models with dynamic batching},
  author={Kim, Y and et al.},
  journal={arXiv preprint arXiv:1909.01953},
  year={2019}
}

@article{radford2018improving,
  title={Improving language understanding by generative pre-training},
  author={Radford, Alec and Narasimhan, Karthik and Salimans, Tim and Sutskever, Ilya and others},
  year={2018},
  publisher={San Francisco, CA, USA}
}

@article{touvron2023llama,
  title={Llama: Open and efficient foundation language models},
  author={Touvron, Hugo and Lavril, Thibaut and Izacard, Gautier and Martinet, Xavier and Lachaux, Marie-Anne and Lacroix, Timoth{\'e}e and Rozi{\`e}re, Baptiste and Goyal, Naman and Hambro, Eric and Azhar, Faisal and others},
  journal={arXiv preprint arXiv:2302.13971},
  year={2023}
}

@article{laguna2022hardware,
  title={Hardware-software co-design of an in-memory transformer network accelerator},
  author={Laguna, Ann Franchesca and Sharifi, Mohammed Mehdi and Kazemi, Arman and Yin, Xunzhao and Niemier, Michael and Hu, X Sharon},
  journal={Frontiers in Electronics},
  volume={3},
  pages={847069},
  year={2022},
  publisher={Frontiers Media SA}
}

@misc{nvidia_agx_orin_power,
  author       = {{NVIDIA}},
  title        = {{Jetson AGX Orin Power Consumption in Deep Sleep Mode and in Another Mode}},
  howpublished = {NVIDIA Developer Forums},
  year         = {2022},
  url          = {https://forums.developer.nvidia.com/t/jetson-agx-orin-power-consumption-in-deep-sleep-mode-and-in-another-mode/214816},
  note         = {Accessed: Oct. 6, 2026}
}

@article{haensch2023compute,
  title={Compute in-memory with non-volatile elements for neural networks: a review from a co-design perspective},
  author={Haensch, Wilfried and Raghunathan, Anand and Roy, Kaushik and Chakrabarti, Bhaswar and Phatak, Charudatta M and Wang, Cheng and Guha, Supratik},
  journal={Advanced Materials},
  volume={35},
  number={37},
  pages={2204944},
  year={2023},
  publisher={Wiley Online Library}
}

@article{gul2023finfet,
  title={FinFET 6T-SRAM all-digital compute-in-memory for artificial intelligence applications: An overview and analysis},
  author={Gul, Waqas and Shams, Maitham and Al-Khalili, Dhamin},
  journal={Micromachines},
  volume={14},
  number={8},
  pages={1535},
  year={2023},
  publisher={MDPI}
}
\end{document}